\documentclass[12pt,a4paper]{article}
\usepackage[utf8]{inputenc}
\usepackage{amsmath}
\usepackage{amsfonts}
\usepackage{amssymb}

\newcommand{\be}{\begin{equation}}
\newcommand{\ee}{\end{equation}}
\newcommand{\beqa}{\begin{eqnarray}}
\newcommand{\eeqa}{\end{eqnarray}}
\newcommand{\nn}{\nonumber}

\newcommand{\comment}[1]{}

\begin{document}
\comment{poznamka na zaciatok : myslim si, ze nam v texte chybaju posunute riadky na zaciatku odstavcov, pridame to? Neni to problem, na 10 minut praca, len neviem aky je JMP vkus}
\begin{center} {\Large\bf Laplace-Runge-Lenz vector in quantum mechanics in noncommutative space} \end{center} \vskip0.5cm

\begin{center} Veronika G\'alikov\'a, Samuel Kov\'a\v cik, Peter Pre\v snajder \end{center}

\begin{center} {\it Faculty of Mathematics, Physics and Informatics,\\ Comenius University of Bratislava, Mlynsk\'a dolina F2, Bratislava, Slovakia} \end{center} \vskip0.5cm

\begin{abstract}
The main point of this paper is to examine a \,"hidden"\, dynamical symmetry connected with the conservation of Laplace-Runge-Lenz vector (LRL) in the hydrogen atom problem solved by means of non-commutative quantum mechanics (NCQM).\\
 The basic features of NCQM will be introduced to the reader, the key one being the fact that the notion of a point, or a zero distance in the considered configuration space, is abandoned and replaced with a \,"fuzzy"\, structure in such a way that the rotational invariance is preserved.\\
 The main facts about the conservation of LRL vector in both classical and quantum theory will be reviewed.\\
 Finally we will search for an analogy in the NCQM, provide our results and their comparison with the QM predictions.\\
 The key notions we are going to deal with are non-commutative space, Coulomb-Kepler problem and symmetry.

\end{abstract}

\section{Introduction}

The main goal we are after is to investigate the existence of dynamical symmetry of the Coulomb-Kepler problem in the quantum mechanics in the  non-commutative space and possibly to find the generalization of the LRL vector for this case.

The Coulomb-Kepler problem corresponding to the motion of a particle in a field of a central force proportional to the inverse square of the distance $(r^{-2})$, was one of the main issues which stood in the centre  of attention at the very beginning of the modern physics.  Newton equation of motion for a particle of mass $m$  is
\be \label{CKP} m\,\dot{\vec{v}}\ =\ -\ q\,\frac{\vec{r}}{r^3}\,. \ee
Here $q$ is just a constant which has to do with the magnitude of the force applied. The system with a central force definitely \textit{is} symmetric, at least the rotational invariance is striking; and the  orbital momentum
\be\label{LRLclas} \vec{L}\ =\ m\,\vec{r}\,\times\,\vec{v} \,\ee
is conserved in any central field.
However, it is in general advantageous to be on lookout for\textit{ all} possible symmetries; and there are indeed more of them present in this case due to the fact that not only is the force central, but in addition has the inverse square dependence of the distance.
It turns out that such a system is in a certain way equivalent to a harmonic oscillator moving in the four dimensional space $\mathbb{R}^4 \simeq \mathbb{C}^2$, which possesses eight integrals of motion. Fixing particularly two them one obtains system equivalent to Coulomb-Kepler problem  (\ref{CKP}). Thus, we should be able to find six integrals of motion. Three have been introduced already, namely the components of $\vec{L}$, the remaining three form the constant vector
\be \vec{A}\ =\ \vec{L}\,\times\,\vec{v}\ + q \,\frac{\vec{r}}{r} \ee
which  is defined only up to an inessential multiplicative constant. Below we will use its normalization as indicated above.

Since gravity was the first interaction people dealt with in modern physics, and because according to Newton the force keeping the solar system together goes as $r^{-2}$, it is natural that the first interpretation of the conserved quantity was provided by astronomy. Just briefly: The vector pointing towards the perihelion of the orbit is conserved; the fixed direction tells us that the perihelion does not precess, while the constant magnitude means that the eccentricity does not change. We know it is not quite like that in our solar system; but that is due to the fact that the force does not exactly meet the requirement of the $r^{-2}$ dependence, mainly due to 
the influence of other planets. This corrections have been estimated, and a tiny difference with the observed precession of Mercury's perihelion was explained by general relativity.
\\

Now we can move from  the classical case  to the QM, or "classical quantum mechanics" (the rather strange combination of words appeared merely to  point out that NCQM, some "generalization of quantum mechanics" will play the prime role throughout this paper).\\
The following few remarks on the subject have been well known  since 1926, when  Wolfgang Pauli published his paper on the subject, \cite{Pa}.  He used the  LRL vector to find the spectrum of hydrogen atom using modern quantum mechanics and the hidden dynamical symmetry of the problem, without the knowledge of the explicit solution of  Schr\"odinger equation. This is not an attempt to provide a full review; if there is a wish for deeper insight, the reader  is invited to read Pauli's original paper.

Coulomb and gravitational forces are so  much alike that it was quite natural to look for deeper analogies, the hydrogen atom being the first thing to begin with. Of course due to the need of QM treatment it is not appropriate to speak of orbits or perihelia, but still. The motivation is quite straightforward - there are two three-component vectors (angular momentum $\vec{L}$ and the LRL vector $\vec{A}$) conserved for a planet orbiting the Sun, we can try to find the quantum mechanical analogs for electron "orbiting" the proton. \\

It turned out that the LRL vector can be found among the hermitian operators acting in the considered Hilbert space in an almost complete analogy with the classical case. One subtlety occurs (due to the ordering issues one often encounters when dealing with non-commuting objects) - the cross product needs to be properly symmetrized, resulting into

 \be
A_k^{QM} = \frac{1}{2}\varepsilon_{ijk} (L_i v_j+ v_j L_i ) + q\frac{x_k}{r},
\ee
where $v_j = - \frac{i\hbar}{m}\, \partial_j $ stands for velocity operator. Vital information  is that the operators $L_i$ and $A_i$ commute with  the Hamiltonian, i.e. are conserved with respect to the  time evolution, and as to their mutual commutation relations, there would be pretty good views of them forming a closed algebra of $so(4)$, if it were not for the commutator $[A_i,A_j] \propto L_k H$.
That commutator depends on the Hamiltonian which was not supposed to be among the algebra constituents; but still we do not have to give up on finding the symmetry. The above mentioned failure to close the algebra simply means that they do not satisfy the $ so(4)$ commutation relations by definition on the whole Hilbert space $\mathcal{H}$. It was the Hamiltonian that spoiled the  commutator  $[A_i,A_j] $; if there is a subspace $\mathcal{H}_E$ spanned by the eigenvectors of $H$ corresponding to the eigenvalue E, then restricting ourselves to $\mathcal{H}_E$ we can replace $H$ with its eigenvalue, which is a c-number. Besides enabling the algebra to close (with $A_i$ being rescaled properly), we  have also dragged the energy into the very definition of the algebra generators. This, together with the theory related to the relevant Casimir operators, has a direct impact on the energy spectrum.\\

To sum up, the components of $\vec{L}^{QM}$ and $\vec{A}^{QM}$  form a representation of generators of a dynamical group on the space  $\mathcal{H}_E$. There are many subspaces $\mathcal{H}_E$ for any admissible eigenvalue $H$ for which $\vec{L}^{QM}$ and $\vec{A}^{QM}$ are given as hermitian operators, i.e., $\mathcal{H}_E$ is a carrier space of unitary representation of the dynamical group. As to the classification of these representations, the Casimir operators are of vital importance. Here we will leave out the theory behind, since the Casimir operators and their eigenvalues for all dynamical groups in question  are well-known. For example, for the Coulomb attractive potential and negative energies the symmetry group is $SO(4)$, its Casimir operators possess integer eigenvalues that are related to energy through the generators; so the discrete energy spectrum is to be expected. In fact Pauli really worked it out and obtained the correct formulas for the hydrogen atom spectrum even before Schr\"{o}dinger.\\

A little historical note should be put here. We call the mentioned quantity Laplace-Runge-Lenz vector, as it is commonly done nowadays. This is, however,  more for the sake of convention than the historical correctness. The vector had been forgotten and brought to light again several times and neither of the three gentlemen was the first one to think of it. We have to go back into the earlier days of physics.

No such quantity was  found among Newton's publications, so he might really not have been aware of the related conservation law.  As far as we know, the first ones to make a mention of it were  Jakob Hermann and  Johann Bernoulli in the letters they exchanged in 1710, see \cite{He}, \cite{Be}. So the name "Hermann-Bernoulli vector" would be probably far more just. It was much later in  1799 that the vector was rediscovered by Laplace  in his Celestial mechanics \cite{La}. Then it appeared as an example  in a popular German textbook on vectors by C. Runge \cite{Ru}, which was referenced by W. Lenz in his paper on the (old) quantum mechanical treatment of the Kepler problem or hydrogen atom \cite{Le}.  After Pauli's publication, it became known mainly as the Runge-Lenz vector, nowadays this vector usually carries the name which we use - with certain objections - also in this paper.
\\

We would like to examine whether, and if so then how, is this affected by the non commutativity of the space; this is what this paper is about.

The paper is organized as follows. In Section 2 we present a brief introduction to the quantum mechanics in a spherically symmetric noncommutative space (NC QM) - the Hilbert space of wave functions in NC space and the NC generalizations of important operators: the Hamiltonian, the angular momentum, the coordinate and the velocity operators are to be presented. Using those, there is a proper generalization of dynamical symmetry of the Coulomb-Kepler problem in NC QM presented in Section 3.  The last Section 4 contains conclusions. Detailed calculations are collected in the Appendix.

\section{Basics of noncommutative quantum \\ mechanics}
This section is by no means to be considered an exhaustive introduction to NCQM; it is meant just to provide a brief review.\\ 
Let us just say that the main issue is to introduce some non-trivial structure into the configuration space, so that the notion of a single point has to be given up. The ideas of this kind have been appearing since the eighties of the previous century, see for instance \cite{a}, \cite{b}. The original motivation was probably the hope of removing divergences from quantum field theories, and maybe of building a base for quantum gravity. However, these tasks have turned out to be more difficult than expected. \\

Learning the lesson from those attempts, we prefer to start small.
So although the  primary ambition of NCQM has never been the "improvement" of  the precision of standard QM, one has to start somewhere; and it is at least safer to introduce the concept of NC space while dealing with QM problems and test it there before handling any of the QFT issues.\\

The first thing which should be made clear is how do we introduce some uncertainty principle into the configuration space for the Coulomb problem without spoiling the key feature which allows us to find the exact solution - the rotational invariance.\\
The uncertainty will be expressed as non-trivial commutation relations for the NC analogs of the former Cartesian coordinates (obviously they cannot  be c-numbers anymore), defined in a spherically symmetrical way.\\

The coordinates in the \textbf{NC configuration space}
${\bf R}^3_\lambda $ are realized in terms of 2 pairs of boson annihilation and creation
operators $a_\alpha$, $a^+_\alpha$, $\alpha\,=\,1,2$,
satisfying the following commutation relations :
\be \label{ca} [a_\alpha,a^+_\beta]\,=\,\delta_{\alpha\beta },\ \ \
[a_\alpha,a_\beta]\,=\,[a^+_\alpha, a^+_\beta]\,=\,0\, .\ee
They act in an \textbf{auxiliary Fock space }${\cal F}$ spanned by normalized
vectors
\be\label{funct} |n_1,n_2\rangle\ =\ \frac{(a^+_1)^{n_1}\,(a^+_2)^{n_2}}{
\sqrt{n_1!\,n_2!}}\ |0\rangle\,. \ee
The normalized vacuum state is denoted as  $|0\rangle\,\equiv\,|0,0\rangle$ We shall write ${\cal F}_n\, =\, \{ |n_1,n_2\rangle\,|\ n_1+n_2 = n\}$.
\\
The \textbf{noncommutative coordinates $x_j$}, $j\,=\,1,2,3$ are given as
\be\label{sph3} \ x_j\ =\ \lambda\,a^+\,\sigma^j\,a\ \equiv\
\lambda\,\sigma^j_{\alpha\beta}\,a^+_\alpha\,a_\beta,\
j\,=\,1,2,3\,,\ee
where  $\sigma^j $ are the Pauli matrices and $\lambda$ is a length parameter, with a magnitude that is not fixed within our model; it is related to the smallest distance relevant in the given non-commutative configuration space  denoted as ${\bf R}^3_\lambda $. We just know $\lambda$  is small enough to avoid any experimental evidence so far. It may well be even at the Planck's scale; if it is so, then the NC corrections are hidden beyond the scale of ordinary QM experiments. \\
The NC analog of the Euclidean \textbf{distance from the origin} is
$r\,=\,\lambda\,(N + 1)$, $N = a^+_\alpha a_\alpha$.
The key  rotationally invariant relations in the theory are
\be\label{ncx} [x_i,x_j]\ =\ 2i\,\lambda\,\varepsilon_{ijk}\,x_k\,,
\ \ \ \ [x_i,r]\,=\,0\,,\ \ \ \ r^2 - x_j^2\,=\, \lambda^2\,.\ee
The first equation defines non-commutative or fuzzy sphere that appeared long time ago in various contexts \cite{berezin} - \cite{GrPr}, e.g., quantization on a sphere = non-flat phase space, a simple model of NC manifolds. In \cite{GrKoPr}  a QFT model on fuzzy sphere  realized in terms of two pairs of annihilation and creation operators was proposed. All these models were considering a single fuzzy sphere.

Here we shall deal with an infinite sequence of fuzzy spheres  dynamically via NC analog of the (radial) Schr\"odinger equation that will be introduced below.\\

Constructing NC quantum mechanics we have first to decide on a  \textbf{Hilbert space ${\cal H}_\lambda $} of states (see also \cite{us}). The suitable choice is a linear space of normal ordered analytic functions containing the same number of creation and annihilation operators:
\be\label{Hilb} \Psi\ =\ \sum\, C_{m_1 m_2 n_1 n_2}\,(a^+_1
)^{m_1}\,(a^+_2)^{m_2}\,(a_1)^{n_1}\,(a_2)^{n_2}\,,\ee
which possess finite weighted Hilbert-Schmidt norm
\be\label{whs1} \| \Psi \|^2\ =\ 4\pi\,\lambda^3\,\mbox{Tr}
[(N+1)\,\Psi^\dagger\,\Psi]\ =\ 4\pi\,
\lambda^2\,\mbox{Tr}[r\,\Psi^\dagger\,\Psi] \,.\ee
(the summation in (\ref{Hilb}) is over nonnegative integers satisfying
$m_1+m_2 \,=\,n_1+n_2$.)\\

Of course our NC wave functions $\Psi$  are themselves operators on the Fock space already mentioned. This does not need to confuse us; the relations which will occur  are to be looked at as operator equalities - we may write $ |n_1,n_2\rangle\ $  on both sides on every such equation. This may be just a formality sometimes. There are cases, however, when it shortens the calculations.

We have got a bit tricky situation here, since the NC coordinates and the NC wave functions (which are to be viewed as elements of the Hilbert space of states), are composed of operators on the auxiliary Fock space. Now, we want to define operators acting on the Hilbert space.  To confuse the reader as little as possible, we have decided from now on to leave the NC coordinates and the NC wave functions $\Psi$ (operators on the Fock space) as they are,  and to denote the operators acting on $\Psi$ with a hat.
\\
\\
The generators of rotations in ${\cal H}_\lambda $: \textbf{orbital momentum operators}
 are defined as
\be\label{L0} \hat{L}_j\,\Psi\ =\ \frac{1}{2}\,[a^+\,\sigma^j\,a,\Psi] \, = \frac{1}{2\lambda}\,(x_j,\Psi -\Psi x_j),\,\, \ j\,=\,1,2,3\,.\ee
They are hermitian and obey the usual commutation relations
\be [\hat{L}_i,\hat{L}_j] \Psi\,\equiv\,(\hat{L}_i \hat{L}_j\,-\,\hat{L}_j \hat{L}_i)
\Psi\,=\, i\,\varepsilon_{ijk}\, \hat{L}_k \Psi\,.\ee
\\
The standard eigenfunctions $\Psi_{jm}$, $j =
0,1,2,\,\dots,\,$ $m = -j,\,\dots,\,+j$, sa\-tisfying
\be\label{Hjm1} \hat{L}_i^2\,\Psi_{jm}\ =\ j(j+1)\,\Psi_{jm},\ \ \
\hat{L}_3\,\Psi_{jm}\ =\ m\,\Psi_{jm}\ ,\ee
are given by the formula
\be\label{harm2} \Psi_{jm}\ =\ \sum_{(jm)}\
\frac{(a^+_1)^{m_1}\,(a^+_2)^{m_2}}{m_1!\,m_2!}\ R_j(r
)\ \frac{a^{n_1}_1\,(-a_2)^{n_2}}{n_1!\ n_2!}\,, \ee
where the summation goes over all nonnegative integers satisfying
\[ m_1+m_2 \,=\,n_1+n_2\,=\,j\,,\ \ \  m_1-m_2-n_1+n_2\,=\,2 m\,. \]
For any fixed $R_j(r)$ equation (\ref{harm2}) defines a representation space for a unitary irreducible
representation with spin $j$. \\
\\

The NC analog of the usual \textbf{Laplace operator }is
\be\label{Lapl} \hat{\Delta}_\lambda\,\Psi\ =\ -\,\frac{1}{\lambda r}\
[\hat{a}^+_\alpha,\,[\hat{a}_\alpha,\,\Psi]]\ =\ -\,\frac{1}{\lambda^2
(N+1)}\ [\hat{a}^+_\alpha ,\,[\hat{a}_\alpha ,\,\Psi]]\ .\ee
\\
\\
The operator $\hat{U}$ corresponding to a\textbf{ central potential} in QM is defined simply as the
multiplication of the NC wave function by $U(r)$:
\be (\hat{U} \Psi)(r)\ =\ U(r)\,\Psi \ =\ \Psi\,U(r)\,.\ee
Since any term of $\Psi\,\in\,{\cal H}_\lambda $ contains the same number of
creation and annihilation operators (any commutator of such a term with $r$ is zero), the left and right multiplications by $U(r)$ are equal.
\\

Now we can define our \textbf{NC Hamiltonian}:
\be\label{NCHamiltonian} \hat{H} \Psi\ =\ \frac{\hbar^2}{2m\lambda r}\ [\hat{a}^+_\alpha ,\,
[\hat{a}_\alpha ,\,\Psi]]\ -\ \frac{q}{r}\, \Psi \ .\ee
\\
As to the \textbf{velocity operator}, it should be related to the evolution of coordinate operator. Let us not worry about the fact that time was not introduced into this theory, the evolution is given by the Hamiltonian. The NC analog of the time derivative is proportional to the commutator of the considered quantity with $H$; so the components of the velocity operator are computed as 
\be\label{V_j}
\hat{V}_j \Psi = -i [\hat{x}_j, \hat{H}]\Psi
\ee
The coordinate operator $\hat{x}_j$ acts on $\Psi$ symmetrically as $\hat{x}_j\Psi= 1/2 (x_j\Psi + \Psi x_j)$. Both sets of NC observables, $\hat{V}_j $ and  $\hat{x}_j$, have been introduced in \cite{SKPP}. Below, we shall see that they are well adapted to the construction of NC analog of the LRL vector.
\\
\\
Based on what has been briefly summed up above, the \textbf{NC analog of the
Schr\"odinger equation} with the Coulomb potential in ${\bf R}^3_\lambda$ is postulated:
\be\label{ncsch1} \frac{\hbar^2}{2m\lambda r}\,[\hat{a}^+_\alpha ,
[\hat{a}_\alpha ,\Psi]] - \frac{q}{r}\, \Psi = E\,\Psi
\ee
To disburden the expressions already crowded enough, from now on we will usually set $m=1$, $\hbar =1$ .

We have been dealing with the above mentioned Schr\"odinger equation in more detail in another paper \cite{us}, where the solution has been found in terms of hypergeometric functions. The energy spectrum has been examined, too. NC corrections to the standard QM prediction have been found, all vanishing in the \, "commutative limit" $\lambda\rightarrow 0$. In this paper we  are going to adopt a different approach - we will try to derive the spectrum using the higher symmetry which the problem  seems to have. There will be an assumption that the eigenfunctions of the Hamiltonian in (\ref{ncsch1}) exist, but we will not need to know their explicit form. It will be interesting to compare the predictions with the results we have got in \cite{us}.

\section{Dynamical symmetry in NCQM}
We shall move on to the NCQM version of the Coulomb-Kepler problem now. The question is, whether we can find sensible analogs of the three components $A_i$ of the LRL vector in such a way, that  all the requirements regarding commutation relations are satisfied (the commutator with the Hamiltonian has to be zero because of the conservation law and relations among all components of $\vec{A}$ and $\vec{L}$ are supposed to correspond to the relevant symmetry).
\\

We are going to answer this question by actually finding the NC version of $A_i$. There is no prescribed way to launch the search for it - in this case it was done by a combination of an educated guess and a piece of good luck. The "educated" part involves trying to keeping the procedure  as close to the classical case as possible. To build up the "original" classical  LRL vector one just has to suitably combine the components of velocity, angular momentum and position vectors (it is no big deal  to replace momentum by velocity multiplied by mass in the classical mechanics, but an important step to introduce the idea that we actually have the NC analogs of all that is needed and the task reduces to finding the proper way of combining it together).

Recall the lesson taken from QM: When constructing the QM version of LRL vector, the cross product of momentum and angular momentum needed to be symmetrized due to their non-vanishing commutator. The NC operators we are going to use when constructing the analog of the cross product part, i.e. $\hat{V}_i$, $\hat{L}_i$ do not commute either, so some symmetrization of the mentioned sort is supposed to take place as well.   Yet this is not all,  an additional subtlety  is to be taken into account: there is another, "potential" part of the LRL vector, in both classical and  quantum mechanics proportional to $ \vec{r}/ r$.  In QM $x_i/r$ was supposed to just multiply the wave function, but since the corresponding NC analogs of $x_i$ and $\Psi$ do not commute, the ordering in the product makes a difference and there is no reason to prefer either of the two possibilities.  We resolve this the same way we did in the cross product case - we choose both and take the average:

\be\label{A}
\hat{A}_k\ =\ \frac{1}{2}\,\varepsilon_{ijk}\, (\hat{L}_i \hat{V}_j\, +\, \hat{V}_j \hat{L}_i)\, +\, q\, \frac{\hat{x}_k}{r}\ ,
\ee
where $\hat{x}_k$ acts as $\hat{x}_k \Psi = 1/2 (x_k \Psi + \Psi x_k)$; recall the definition of the velocity operator.\\

Now comes the  "lucky" part of the story - besides coping with the ordering  dilemma, nothing  more needs to be done, except for actually working out the calculations to justify our definition of $\vec{A}$. That requires more work than  the  previous sentence may suggest, and the better part of this paper deals with it.

So now we are going to take the NC analogs of the Hamiltonian, velocity, angular momentum and position operators, mix them together according to the recipe similar to that known from the classical QM case, and symmetrize what should be symmetrized.

Then the main work will follow - evaluating the commutator $[\hat{A}_i,\hat{H}]$, examining the commutation relations between $\vec{A}$ and $\vec{L}$, rescaling $\vec{A}$ by a suitable numerical factor if needed... all in all,  searching for the signs of a higher dynamical symmetry. Once the symmetry group is known to be present, we can construct the corresponding Casimir operators. We know  how Casimirs for $so(4)$ look like, and their prescribed eigenvalues  are "responsible" for the discrete energy spectrum.

All those that play  important roles: the Hamiltonian, velocity, angular momentum and position operators, have been defined already in terms of creation and annihilation operators $a^+_\alpha$, $ a_\alpha$;  we know commutation relations for these, so we should be able to calculate all that is needed. However, after writing it all down and trying to make heads and tails of it, one quickly comes to the conclusion that the problem is not assigned in the most friendly way. It is better to think twice before introducing any new symbols, but this is definitely the case when it may help. There are certain combinations of $a^+_\alpha$, $ a_\alpha$ that our expressions have in common, so separating them the right way makes the calculations easier.

\subsection{Auxiliary operators}

We are interested in the way in which the considered operators act on the wave functions $\Psi$.
They are expressed in terms of $a^{+}_\alpha$, $ \hat{a}_\alpha$. Generally it matters whether the creation and annihilation operators act from the right or the left and the following notation will turn out to be useful.
\begin{eqnarray}
\hat{a}_\alpha \,\Psi = a_\alpha\, \Psi\,, &\,&\hat{b}_{\alpha }\,\Psi = \Psi\, a_{\alpha}\,, \\
\hat{a}^+_\alpha\, \Psi = a^+_\alpha\, \Psi\,, &\,&\hat{b}^{+}_{\alpha }\,\Psi = \Psi\, a^{+}_{\alpha }\,.
\end{eqnarray}
One obvious virtue of this notation is the fact that from now on we do not have to drag $\Psi$ into the formulas just to make clear which side do the operators act from. The relevant commutation relations are (see \eqref{ca})
\begin{equation} \label{kom}
[\hat{a}_\alpha,\, \hat{a}^+_\beta]\ =\ \delta_{\alpha \beta}\,,\ \ \
[\hat{b}_\alpha,\, \hat{b}^+_\beta]\ =\ -\,\delta_{\alpha \beta}\,.
\end{equation}
The other commutators are zero. Having this last sentence in mind spares a lot of paper while doing the calculations.

As already mentioned, we will use the position operator in the form

\begin{eqnarray} \label{xbar}
 \nonumber \hat{x}_i\,\Psi &=& \frac{1}{2}\,(x_i\, \Psi\, +\, \Psi\, x_i)\,=\, \frac{\lambda}{2}\,\sigma^i_{\alpha \beta}\,(\hat{a}^+_\alpha\, \hat{a}_\beta \, +\, \hat{b}_\beta\, \hat{b}^+_\alpha )\, \Psi\,,\\
  \hat{r}\,\Psi &=&\frac{1}{2}\,(r\, \Psi\, +\, \Psi\, r)\,=\,\frac{\lambda}{2}\,((\hat{a}^+_\alpha\, \hat{a}_\alpha +1)\,  +\, (\hat{b}_\alpha\, \hat{b}^+_\alpha\, +\,1) )\, \Psi \,.
\end{eqnarray}

For the sake of certain future calculations let us also  define
$$
\begin{array}{lcl}
\hat{x}_i^L\, \Psi\, =\, x_i\, \Psi\,,  & \hat{x}_i^R\, \Psi\, =\,  \Psi\, x_i\,, & \hat{x}_i\, =\, \frac{1}{2}\,(x_i^L\, +\, x_i^R)\,, \\
\ \hat{r}^L\, \Psi\, =\, r\, \Psi\,,  & \hat{r}^R\, \Psi\, =\,  \Psi\, r\,, & \  \hat{r} \,=\, \frac{1}{2}\, (r^L\, +\, \hat{r}^R)\,. \\
\end{array}
$$
Note that $\hat{L}_i\, =\, \frac{1}{2\lambda}\,(\hat{x}_i^L-\hat{x}_i^R)$. Functions of $\hat{r}$ are defined in terms of their Taylor expansion $\hat{f}(\hat{r})\, =\, \sum a_n \hat{r}^n$.

 Clearly $\hat{x}_i \neq \hat{x}^{L,R}_i$. On the other hand, the hat over $r$ can be left out if $\hat{r}$ is supposed to act on something that contains  the same number of creation and annihilation operators (for instance all $\Psi$ corresponding to QM $\Psi(\vec{x})$ have this property, see \eqref{sph3}). However, we sometimes keep the $\hat{r}$ symbol just to have the formulas more symmetric. \\

The following sequences of operators will occur so often and their
role is going to be so important that they deserve to have notation on their own:
$$
\begin{array}{lllclll}
\hat{w}_{\alpha \beta} &=& \hat{a}^+_\alpha \hat{b}_\beta - \hat{a}_\beta \hat{b}^+_\alpha\,,  &\,\,\,\, & \hat{\zeta}_{\alpha \beta} &=& \hat{a}^+_\alpha \hat{b}_\beta + \hat{a}_\beta \hat{b}^+_\alpha\,, \\
\hat{w} &=& \hat{w}_{\alpha \alpha} \,,                                    &\,\,\,\, & \hat{\zeta} &= &\hat{\zeta} _{\alpha \alpha} \,,   \\
\hat{w}_k  &=&  \sigma^k_{\alpha \beta}  \hat{w}_{\alpha \beta}\,,           & \,\,\,\,&  \hat{\zeta} _k &=& \sigma^k_{\alpha \beta} \hat{\zeta}_{\alpha \beta} \,.   \\
\end{array}
$$

\begin{eqnarray}
\label{W_k}\hat{W}_k &=&\frac{2\hat{x}_k }{\lambda}-\hat{\zeta} _k  \,\,= \,\, \sigma^k_{\alpha \beta} (\hat{a}^+_\alpha \hat{a}_\beta - \hat{a}^+_\alpha \hat{b}_\beta -\hat{a}_\beta \hat{b}^+_\alpha + \hat{b}^+_\alpha \hat{b}_\beta)\,, \nonumber \\
\hat{W} &=& \frac{2\hat{r}}{\lambda}-\hat{\zeta}   \,\, = \,\, \hat{a}^+_\alpha \hat{a}_\alpha - \hat{a}^+_\alpha \hat{b}_\alpha - \hat{b}^+_\alpha \hat{a}_\alpha +\hat{b}^+_\alpha \hat{b}_\alpha \,,\nonumber\\
&& \nonumber \\
\hat{W}'_k &=& \hat{W}_k + \omega \hat{x}_k =\eta \hat{x}_k - \hat{\zeta}_k \,,\nonumber \\
\hat{W}'& =& \hat{W} + \omega \hat{r} = \eta \hat{r} - \hat{\zeta}\,. \nonumber \\
\label{W} &&
\end{eqnarray}

\noindent $ \sigma^k_{\alpha \beta}$ denotes the Pauli matrices and the  new letters which appeared in the  last two lines are just shorthands:

 $$
\begin{array}{lcl}
\omega\, =\, -2 \lambda E \,,& \mbox{\,\,\,\,}& \eta\, =\, ( \frac{2}{\lambda}\, +\, \omega)\,. \\
  \end{array}
$$
 $E$ is energy and $\lambda$ is the NC parameter mentioned in the introduction. Note that the only difference between   $\hat{W}'_k$  and $\hat{W}_k$ is the constant multiplying one of their terms. $\hat{W}'$  and $\hat{W}$ are related in the same way.

\subsection{NC operators revisited}
We have introduced  new auxiliary operators which should  make the calculations more manageable, now we will rewrite everything relevant in their terms - the Hamiltonian, the velocity operator and the NC LRL vector.

\begin{eqnarray}
\hat{H} &= &\frac{1}{2\lambda \hat{r}}(\hat{a}^+_\alpha \hat{a}_\alpha + \hat{b}^+_\alpha \hat{b}_\alpha - \hat{a}^+_\alpha \hat{b}_\alpha -  \hat{a}_\alpha \hat{b}^+_\alpha ) -\frac{q}{\hat{r}}\ \nonumber \\
\label{H} &= &\frac{1}{2\lambda \hat{r}}(\frac{2\hat{r}}{\lambda}-\hat{\zeta}) -\frac{q}{\hat{r}} \,\,\,=\,\,\, \frac{1}{2\lambda \hat{r}} \hat{W}   -\frac{q}{\hat{r}} \,, \\
&& \nonumber \\
 \label{V_i} \hat{V}_i &=& -i \left[\hat{x}_i, \hat{H} \right] = \frac{i}{2\hat{r}} \hat{w}_i \,, \\
&& \nonumber \\
\label{A_k} \hat{A}_k &= &\frac{1}{2}\varepsilon_{ijk} (\hat{L}_i \hat{V}_j + \hat{V}_j\hat{L}_i) + q \frac{\hat{x}_k}{\hat{r}} \,\, = \,\, -\frac{1}{2\lambda \hat{r}}(\hat{r}\hat{\zeta} _k - \hat{x}_k \hat{\zeta}) + q \frac{\hat{x}_k}{\hat{r}} \nonumber \\
&=& \frac{1}{2\hat{r}\lambda}(\hat{r}\hat{W}_k - \hat{x}_k \hat{W}) + q \frac{\hat{x}_k}{\hat{r}} \,\,=\,\, \frac{1}{2\hat{r}\lambda}(\hat{r}\hat{W}'_k - \hat{x}_k \hat{W}') + q \frac{\hat{x}_k}{\hat{r}} \nonumber \\
&=& \frac{1}{2\hat{r}\lambda}(\hat{r}\hat{W}'_k - \hat{x}_k (\hat{W}' - 2 \lambda q)) \,.
\end{eqnarray}
Deriving equations (\ref{V_i}) and (\ref{A_k}) involves  somewhat  laborious calculations, many steps have been skipped here. However, all of them can be reconstructed from the  definitions given in the preceding paragraphs. Some more details are given in the Appendix, see (\ref{Vcalc}), (\ref{Acalc}). In the second line of (\ref{A_k}) we have used the equality of $(\hat{r}\hat{W}_k - \hat{x}_k \hat{W})=(\hat{r}\hat{W}'_k - \hat{x}_k \hat{W}')$.\\
\\
Now let us rewrite the NC Schr\"{o}dinger equation in the following way:

\be \label{SchEworkingform}
\left(\frac{1}{2\lambda \hat{r}} \hat{W}   -\frac{q}{\hat{r}} - E \right)\Psi_E =\frac{1}{2\lambda \hat{r}}(\hat{W}' - 2 \lambda q) \Psi_E= 0\,.
\ee
\\
$\Psi_E$ belongs to $\mathcal{H}_{\lambda}^E$, i.e. to the subspace spanned by the eigenvectors of the Hamiltonian.

It makes things easier if we  recognize a zero when  stumbling across one. As for (\ref{SchEworkingform}),  it provides a funny form of zero - the calculations, however,  may get less amusing if this point is overlooked.
\\

Here is how does it pay off: Let us examine $\hat{A}_k |_{\mathcal{H}_{\lambda}^E}$, that is, the LRL vector as it works on the solutions of the Schr\"{o}dinger equation. (And once again: we do not have to know the $\Psi_E$ explicitly, they are just supposed to exist and form a basis for the subspace  $\mathcal{H}_{\lambda}^E$.)

\be\label{Aeigen}
\hat{A}_k |_{\mathcal{H}_{\lambda}^E} = \frac{1}{2\hat{r}\lambda}(\hat{r}\hat{W}'_k - \hat{x}_k \underbrace{(\hat{W}' - 2 \lambda q)}_{\mbox{see Eq. (\ref{SchEworkingform}) }})=\frac{1}{2\lambda}\hat{W}'_k \,.
\ee

It is nice to know the "full form" of $\hat{A}_k$, but at the end of the day we are interested in the actual physical bound states, not in the whole abstract Hilbert space. We did not expect the $SO(4)$ symmetry to show somewhere else. So when dealing with calculations related to conservation of $A_k$, we just need to find out whether the following commutator with the Hamiltonian vanishes.

\begin{eqnarray}
\dot{\hat{W}}'_k &=&  i\left[\hat{H}_0 - \frac{q}{\hat{r}} , \hat{W}'_k\right] = i\left[\frac{1}{2\hat{r}\lambda}\hat{W} - \frac{q}{\hat{r}},\hat{W}'_k\right] \nonumber \\
&= & i\left[\frac{1}{2\hat{r}\lambda}\hat{W}' - \frac{q}{\hat{r}},\hat{W}'_k\right] = i\left[\frac{1}{2\hat{r}\lambda}\hat{W}', \hat{W}'_k\right] - iq\left[\frac{1}{\hat{r}},\hat{W}'_k\right] \nonumber \\
\label{dotA} &= & \frac{i}{2\hat{r}\lambda}\left[\hat{W}', \hat{W}'_k\right] + i \left[\frac{1}{\hat{r}}, \hat{W}'_k\right]\left(\frac{\hat{W}'}{2\lambda}-q\right)=0 \,.
\end{eqnarray}
In the second line, $\hat{W}'$ appears instead of $\hat{W}$. It is a legal step to do, since it does not change the commutator, and it is a sensible step too, since we obtained the (\ref{SchEworkingform})-style zero in the second term in the last line (it vanishes  when acting on vectors from $\mathcal{H}_{\lambda}^E$ and we are not interested in the rest of $\mathcal{H}_{\lambda}$). To prove that the first term  proportional to $[\hat{W}', \hat{W}'_k]$ also does not contribute requires calculations lengthy enough to be placed in the Appendix, see (\ref{W0WK}).\\

The equation above apparently conveys an encouragement to search for the underlying $SO(4)$ symmetry, since the \textbf{LRL vector conservation} makes its components suitable candidates for a half of its generators, the remaining three consisting of the components of the angular momentum.
 So let us check it (for detailed derivation see (\ref{tildWW}) in the Appendix)

 \begin{equation}\label{AiAj}
[\hat{A}_i,\hat{A}_j]  =\frac{1}{4\lambda^2} [\hat{W}'_i, \hat{W}'_j]  = i\frac{\omega}{\lambda}\left(1+\frac{\omega\lambda}{4}\right)\varepsilon_{ijk} \hat{L}_k
\end{equation}
or, to see explicitly the energy dependence ($\omega = - 2E \lambda$):
\begin{equation}
[\hat{A}_i,\hat{A}_j] = i \varepsilon_{ijk} \left(- 2E + \lambda^2 E^2\right) \hat{L}_k
\end{equation}
There is nothing but a constant in the way, as long as we let the operator $[\hat{A}_i,\hat{A}_j]$ act upon the vectors from $\mathcal{H}_{\lambda}^E$ with the energy fixed. Eq. (\ref{AiAj}) and
\begin{equation}\label{LA}
[\hat{L}_i, \hat{L}_j] = i \varepsilon_{ijk} \hat{L}_k, \ \ \ [\hat{L}_i,\hat{A}_j] = i \varepsilon_{ijk} \hat{A}_k
\end{equation}
define  Lie algebra relations corresponding to a particular symmetry group which actual form  depends on the sign of the $E$ dependent factor in (\ref{AiAj}). The relevant relations for $\hat{L}_i$ have been already mentioned, so we just need to check the mixed commutator $[\hat{L}_i, \hat{A}_j]$. This is a long process again, so to not distract our attention from what is going on (besides reshuffling operators here and there), we relocated the calculation to the Appendix, see eq. (\ref{liwj}) and the related ones. \\

 There are three independent cases\\

$\bullet $ $SO(4)$ symmetry: $ -2E + \lambda^2 E^2 > 0\ \Longleftrightarrow \ E < 0 $ or $E > 2/\lambda^2 $,

$\bullet $ $SO(3,1)$ symmetry: $ - 2E + \lambda^2 E^2 < 0\ \Longleftrightarrow \ 0 <  E < 2/\lambda^2 $,

$\bullet $ $E(3)$ Euclidean group:  $ - 2E + \lambda^2 E^2 = 0\ \Longleftrightarrow \ E = 0 $ or $E = 2/\lambda^2 $.\\

The admissible values of $E$ should correspond to the unitary representations of the symmetry in question. This requirement guarantees that the generators $\hat{L}_j$ an  $\hat{A}_j$ are realized as hermitian operators, and consequently correspond to physical observables. The Casimir operators in all mentioned cases are
\begin{eqnarray}
\nn \hat{C}^\prime_1 &=& \hat{L}_j \hat{A}_j\,,\\
\nn\hat{C}^\prime_2 &=& \hat{A}_i\hat{A}_i+(- 2E + \lambda^2 E^2)(\hat{L}_i\hat{L}_i+1) \\
\label{CAS} &=& \frac{1}{4\lambda^2} \left(\hat{W}'_i\hat{W}'_i +(\eta^2\lambda^2-4)(\hat{L}_i\hat{L}_i+1) \right) \,.
\end{eqnarray}
The prime indicates that we are not using the standard normalization of Casimir operators.

Now, we need to calculate their values in $\mathcal{H}_{\lambda}^E$. The first Casimir  is vanishing in all cases due to the fact that $\hat{C}^\prime_1 \Psi_E \sim r \Psi_E - \Psi_E r = 0$. (See (\ref{zeroCas1}), (\ref{zeroCas2}) in the Appendix.) The second Casimir operator is somewhat more demanding, and either believe it or convince yourself (consulting the part of the Appendix beginning with (\ref{id}) may help in the latter case), the terms in the bracket add up  exactly to  $(\hat{W}')^2$. According to the Schr\"{o}dinger equation, $(\hat{W}')^2\Psi_E = 4\lambda^2 q^2 \Psi_E $, and we are left with
\begin{equation} \label{CAS2}
\hat{C}^\prime_2\, \Psi_E \ =\ \left( \hat{A}_i\hat{A}_i\,+\,(- 2E + \lambda^2 E^2) \, (\hat{L}_i\hat{L}_i+1) \right) \Psi_E \ =\ q^2\, \Psi_E \,.
\end{equation}
Since both Casimir operators take  constant values $\hat{C}^\prime_1 = 0$ and $\hat{C}^\prime_2 = q^2$ in $\mathcal{H}_{\lambda}^E$, we are dealing with irreducible representations of the dynamical symmetry group $G$ that are unitary for particular values of energy. In all considered cases, $G = SO(4), SO(3,1), E(3)$, the unitary irreducible representations are well known. The corresponding systems of eigenfunctions that span the representation space have been found in \cite{us}. Here we shall not repeat their construction, but we shall restrict ourselves to brief comments pointing out some interesting aspects.\\

{\bf 1. Bound states -} $\boldsymbol{SO(4)}$ {\bf symmetry:} $\boldsymbol{- 2E + \lambda^2 E^2 > 0}$. In this case we rescale the LRL vector as

\be
\hat{K}_j \ = \frac{\hat{A}_j}{\sqrt{- 2E + \lambda^2 E^2}}\ =\  \frac{\hat{W}'_j}{\sqrt{\eta^2\lambda^2-4}}\,.
\ee
After this step Eqs. (\ref{AiAj}), (\ref{LA}) turn into the following relations:

\begin{equation}\label{lilj}
[\hat{L}_i, \hat{L}_j] = i \varepsilon_{ijk} \hat{L}_k \,,\ \ \
[\hat{L}_i, \hat{K}_j] = i \varepsilon_{ijk} \hat{K}_k \,, \ \ \
[\hat{K}_i, \hat{K}_j] = i \varepsilon_{ijk} \hat{L}_k \,.
\end{equation}
Thus we have got the representation of the $so(4)$ algebra. The relevant normalized Casimir operators read
\begin{equation}
\hat{C}_1\ =\ \hat{L}_j\, \hat{K}_j\,,\ \ \  \hat{C}_2\ =\ \hat{K}_i\,\hat{K}_i\,+\,\hat{L}_i\hat{L}_i\,+\,1\,.
\end{equation}
As we have stated already, the $\hat{C}_1$ acting on an eigenfunction of the Hamiltonian returns zero. As to $\hat{C}_2$, we know that for $so(4)$,  under the condition that the first Casimir is zero, the second one  has to equal to $(2j+1)^2$ for some integer or half-integer $j$. At the same time, according to (\ref{CAS2}) it is related to the energy:

\begin{eqnarray}\label{cas2}
 \hat{K}_i\,\hat{K}_i\,+\,\hat{L}_i\,\hat{L}_i+1 &=&  \frac{4\lambda^2 q^2}{\eta^2\lambda^2-4}\,=\, \frac{ q^2}{\lambda^2 E^2 - 2E}\,, \nonumber \\
 & &    \nonumber \\
(2j+1)^2 &= & \frac{ q^2}{\lambda^2 E^2 - 2E}\ .
\end{eqnarray}
We will write $n^2$, $n= 1,\, 2,\, ...$ instead of $(2j+1)^2 $.  Now solving the quadratic equation for energy  we obtain two discrete sets of solutions depending on $n$:
\begin{equation}\label{spectrum+-}
E\ =\  \frac{1}{\lambda^2}\, \mp\, \frac{1}{\lambda^2} \sqrt{1+\kappa_n^2}\,,\ \ \ \kappa_n\ =\ \frac{q\lambda}{n} \,.
\end{equation}

The first set of eigenfunctions of the Hamiltonian in (\ref{NCHamiltonian}) for energies $E < 0$ (i.e. negative sign in front of the square root in (\ref{spectrum+-})) has been found for Coulomb {\it attractive} potential, i.e. $q>0$ in (\ref{NCHamiltonian}). The eigenvalues possess smooth standard limit for $\lambda\rightarrow 0$:

\begin{eqnarray}\label{spectrum}
E^I_{\lambda n} &=&  \frac{1}{\lambda^2}\, -\, \frac{1}{\lambda^2} \sqrt{1+\kappa_n^2}\, =\,  \frac{1}{\lambda^2}\, -\, \frac{1}{\lambda^2}\, \left(1\,+\, \frac{1}{2}\, \kappa_n^2\, -\, \frac{1}{24}\, \kappa_n^4\, +\, ...\right)\nonumber \\
 & &    \nonumber \\
& =&   -\frac{q^2}{2n^2}\, +\, \lambda^2\, \frac{q^4 }{24 n^4}\, +\, ... \nonumber \\
 & &    \nonumber \\
& &   \mbox{and after restoring the symbols  $m$, $\hbar$} \nonumber \\
 & &    \nonumber \\
E^I_{\lambda n}&= &   -\frac{q^2m}{2n^2\hbar^2}\, +\, \lambda^2 \frac{q^4 m^3}{24 n^4\hbar^6} + ...
\end{eqnarray}
The "..." stands for the terms proportional to higher (even) powers of $\lambda$. This solution is similar to the spectrum for Coulomb {\it attractive} potential, $q>0$, that has been found using algebraic methods by Pauli prior to solving Schr\"odinger equation for the hydrogen atom. In fact the results coincide when we consider the limit $\lambda \rightarrow 0$ in (\ref{spectrum}).

The full set of eigenfunctions of (\ref{NCHamiltonian}) for energies $E < 0$ was constructed in \cite{us} by explicitly solving the NC Schr\"odinger equation. The radial NC wave functions defined in (\ref{harm2}) are given in terms of hypergeometric function
\begin{eqnarray}
\nn R^I_{\lambda n} &=& (\Omega_n)^N\, F(-n,\, -N,\, 2j+2,\, -2 \kappa_n\, \Omega_n^{-1})\,,\\
\label{radI} \Omega_n &=& \frac{\kappa_n - \sqrt{1+\kappa_n^2}+1}{\kappa_n + \sqrt{1+\kappa_n^2} -1}\,,
\end{eqnarray}
where $N = a^+_\alpha a_\alpha$ controls the radial NC variable.

The second set of very unexpected solutions corresponds to energies (\ref{spectrum+-}) with positive sign
\begin{equation}\label{spectrum'}
E^{II}_{\lambda n}\ =\  \frac{1}{\lambda^2}\, + \,\frac{1}{\lambda^2}\, \sqrt{1+\kappa_n^2}\, >\, \frac{2}{\lambda^2} \,.
\end{equation}
The corresponding radial NC wave functions has been found in \cite{us} solving NC Schr\"odinger equation for a Coulomb {\it repulsive}  potential, $q<0$ in (\ref{NCHamiltonian}). These radial NC wave functions are closely related to those given above
\begin{equation}\label{radII}
R^{II}_{\lambda n}\ =\ (-\Omega_n)^N\, F(-n,\, -N,\, 2j+2,\, 2 \kappa_n\, \Omega_n^{-1})\,.
\end{equation}.

Let us denote the first set of eigenfunctions for the Coulomb attractive potential defined by (\ref{harm2}) with radial part $R^I_{\lambda n}$  as $\Psi^I_{nlm}$, and similarly, the second set of solutions with radial part $R^{II}_{\lambda n}$ for Coulomb repulsive potential by $\Psi^{II}_{nlm}$. The mapping
$$ \Psi^I_{nlm}\ \longmapsto\  \Psi^{II}_{nlm} $$
is obviously a unitary transformation in ${\cal H}_\lambda$. Thus both $SO(4)$ representations, the one for Coulomb attractive potential with $E^I_{\lambda n} < 0$ and that for ultra-high energies $E^{II}_{\lambda n} > 2/\lambda^2$ for Coulomb repulsive potential, are unitary equivalent. That is natural, since in both representations the Casimir operators take the same values, $\hat{C}_1 = 0$ and $\hat{C}_1 = n^2$. However, in the commutative limit $\lambda\rightarrow 0$ the extraordinary bound states at ultra-high energies disappear from the Hilbert space.\\

 {\bf 2. Coulomb scattering:} $\boldsymbol{2E - \lambda^2 E^2 > 0$ $\Longleftrightarrow$ $0 < E < 2/\lambda^2}$. In this case we rescale the LRL vector as
\be
\hat{K}_j =\frac{\hat{A}_j}{\sqrt{2E - \lambda^2 E^2}} = \frac{\hat{W}'_j}{\sqrt{4-\eta^2\lambda^2}},
\ee
  After this step we obtain equations
\begin{equation}\label{lilj'}
[\hat{L}_i,\, \hat{L}_j]\, = \,i\, \varepsilon_{ijk}\, \hat{L}_k \,, \ \ \
[\hat{L}_i,\, \hat{K}_j]\, =\, i\, \varepsilon_{ijk}\, \hat{K}_k \,, \ \ \
[\hat{K}_i,\, \hat{K}_j]\, =\, - i\, \varepsilon_{ijk}\, \hat{L}_k \,.
\end{equation}
So this time we have got the representation of the $so(3,1)$ algebra. The relevant normalized Casimir operators read
\begin{equation}
\hat{C}_1\ =\ \hat{L}_j\, \hat{K}_j\,,\ \ \  \hat{C}_2\ =\ \hat{K}_i\,\hat{K}_i \,-\, \hat{L}_i\,\hat{L}_i\,.
\end{equation}

In our case $\hat{C}_1 = 0$, so we are dealing with $SO(3,1)$ unitary representations that are labeled by the value of second Casimir operator $\hat{C}_2 = \tau $, see e.g \cite{BarutRanczka}:

$\bullet $ Spherical principal series for $\tau > 1$;

$\bullet $ Complementary series for $ 0 < \tau < 1 $.\\
Let us point out that for integer $\hat{C}_1 \neq 0$ and arbitrary real $\hat{C}_2$ a (non-spherical) remainder of principal series appears that completes the set of unitary representations of $SO(3,1)$ group.\\

Rewriting (\ref{CAS2}) in terms of $\hat{K}_j$ we obtain relation between energy $E$ and the parameter $\tau$:
\begin{equation}
\hat{K}_i\,\hat{K}_i\,-\,\hat{L}_i\,\hat{L}_i\ =\ 1\, +\, \frac{q^2}{2E - \lambda^2 E^2}\ =\ \tau\ >\ 1\,.
\end{equation}
Thus we are dealing with the principal series $SO(3,1)$ unitary representations. The scattering NC wave functions have been constructed in \cite{us} for any admissible energy $E \in (0,\,2/\lambda^2)$, and from their asymptotic behavior the partial wave $S$-matrix has been derived
\begin{equation}
S^\lambda_j (E)\ =\ \frac{\Gamma (j + 1 - i \frac{q}{p})}{\Gamma (j + 1 + i \frac{q}{p})}\,,\ \ \ p\ =\ \sqrt{2E - \lambda^2 E^2 }\ .
\end{equation}

It can be easily seen that such $S$-matrix possesses poles at energies $E = E^I_{\lambda n}$ for Coulomb attractive potential and  poles at energies $E = E^{II}_{\lambda n}$ for Coulomb repulsive potential, where both $E^I_{\lambda n}$ and $E^{II}_{\lambda n}$ coincide with (\ref{spectrum}) and (\ref{spectrum'}) given above. As for energies
\begin{equation}
E_\mp\ =\ \frac{1}{\lambda^2} \left( 1\, \mp\, \sqrt{1 - \frac{\lambda^2 q^2}{\tau -1}}\right)
\end{equation}
the values of Casimir operators coincide, the corresponding representations are unitary equivalent. This relates the scattering for low energies $0 < E < 1/\lambda^2$ to that at high energies $1/\lambda^2 < E < 2/\lambda^2$.

We skip the limiting cases of the scattering at the edges $E = 0$ and $E = 2/\lambda^2$ of the admissible interval of energies, where the $SO(3,1)$ group contracts to the group $E(3) = SO(3)\triangleright T(3)$ of isometries of 3D space with generators $\hat{L}_j$ and $\hat{L}_j$ satisfying commutation relations (see \cite{BarutRanczka}):
\begin{equation}
[\hat{L}_i,\, \hat{L}_j]\ =\ i\, \varepsilon_{ijk}\, \hat{L}_k\,, \ \ \ [\hat{L}_i,\,\hat{A}_j]\, =\, i\, \varepsilon_{ijk}\, \hat{A}_k\,, \ \ \ [\hat{A}_i,\,\hat{A}_j]\, =\,0\,.
\end{equation}
The corresponding NC hamiltonian eigenstates are given in \cite{us}.

\section{Conclusions}
This paper is focused on the Coulomb-Kepler problem in non-commutative space.
We have found the NC analog of the LRL vector; it's components,   together with those of  the NC angular momentum operator,  supply the algebra of generators of a symmetry group. It is remarkable that the  formula for the NC generalization of LRL vector looks like a carbon copy of the standard formula when written in terms of the proper NC observables: NC angular momentum, NC velocity, symmetrized NC coordinate and NC radial distance.\\

The group $SO(4)$ has appeared twice: Firstly,   we have been dealing with bound states for negative energies in the case of the attractive Coulomb potential, which have an analog in the standard quantum mechanics; and secondly,   we have found an unexpected set of bound states for positive energies above certain ultra-high value in the case the potential is repulsive.\\

$SO(3,1)$ is the symmetry group to be considered when examining the scattering (relevant for the interval of energies between zero and the mentioned critical ultra-high  value).\\

The conservation  of our NC LRL vector  has been shown, as well as the commutation relations  related to $so(4)$ and $so(3,1)$ algebra. In case of the bound states (of both kinds), the calculations of the relevant Casimir operators  have revealed that there are just specific values that the energy can take. The explicit formulas for the energy spectra have been given. The result for negative energy bound states  coincides with the well known QM prediction in the \,"commutative" limit  $\lambda \rightarrow  0 $.  Besides that, the NC corrections proportional to $\lambda^2$ and higher (even) powers occur. As to the unexpected ultra-high energy bound states, they disappear from the Hilbert space providing that $\lambda \rightarrow  0 $.  \\

So now we have NCQM predictions for  the hydrogen atom energy spectrum obtained in two ways:\\
- the results  given by  in this paper,  obtained using the dynamical symmetry in analogy with Pauli's method  (without explicitly solving the NC Schr\"{o}dinger equation for eigenstates of the Hamiltonian)\\
- the result published in \cite{us},  acquired by explicitly solving  the NC analog of the Schr\"{o}dinger equation  (without using those dynamical symmetry arguments). \\

We are glad to find out that these outcomes \textit{do }agree.

\section{Appendix}
Here we provide some more detailed calculations that have been skipped in order to avoid overloading the previous sections with profusion of technical details which would hardly have been helpful in keeping track of the main ideas.\\
Even here we are not about to write down every single step; if any interest in tiny details arises, we suppose it will be easier for the reader to calculate something on their own now and then, than to try to keep track of all the signs and many times renamed indices in this paper. Therefore we provide the main tricks which make the calculations more manageable, and then we write down the sketches of the particular derivations.\\
\\
We can almost say that everything one needs to do is to calculate commutators of various strings of creation and annihilation operators acting from the left or from the right, that is, using the commutation relations for $\hat{a}^+_\alpha ,\,\hat{a}_\beta, \,\hat{b}^+_\alpha ,\,\hat{b}_\beta $ in a suitable way. As we know already, the $b$- operators are simply the $a$-ones acting from the right; the fact that those acting from the opposite sides commute saves us a good deal of work. Besides that, one should be familiar with various identities which hold for the Pauli matrices.\\
\\
So the following should be  borne in mind before digging oneself into the calculations:
$$
\begin{array}{lllclll}
 [\hat{a}_\alpha, \hat{a}^+_\beta] &=& \delta_{\alpha \beta}&\,\,\,\,\,\, &[\hat{b}_\alpha, \hat{b}^+_\beta] &=&-\delta_{\alpha \beta} \\
\end{array}
$$
The other commutators are zero.\\

Below we shall use frequently the identities for Pauli matrices a the identities for NC coordinates that follow directly:

$$
\begin{array}{lllclll}
[\sigma^i,\sigma^j] &=&2i\varepsilon_{ijk} \sigma^k &\,\,\,\, & \varepsilon_{ijk} \sigma^i_{\alpha\beta}\sigma^j_{\gamma\delta} &=&i(\sigma^k_{\alpha\delta}\delta_{\gamma\beta}-\sigma^k_{\gamma\beta}\delta_{\alpha\delta}) \\
 \{\sigma^i,\sigma^j\} &=&2\delta_{ij}\mathbf{1}                                 &\,\,\,\, &  \sigma^i_{\alpha\beta}\sigma^i_{\gamma\delta}  &=&2\delta_{\alpha\delta}\delta_{\gamma\beta}- \delta_{\alpha\beta}\delta_{\gamma\delta}\\
\sigma^i\sigma^j &=&\delta_{ij}\mathbf{1} +i\varepsilon_{ijk} \sigma^k             & \,\,\,\,&  Tr\sigma^i &=&\sigma^i_{\alpha\alpha} = 0   \\

 &&&& \\

 [\hat{x}_i,\hat{r}] &=& 0  &\,\,\,\, & [\hat{x}_i^L, \hat{x}_j^R] &=&0 \\

[\hat{x}_i^L, \hat{x}_j^L] &=& 2i\lambda\varepsilon_{ijk}  \hat{x}_k^L      &\,\,\,\, &   [\hat{x}_i^R, \hat{x}_j^R] &=& -\,2i\lambda\varepsilon_{ijk}  \hat{x}_k^R\\

\end{array}
$$
Make sure  you have noticed the minus sign in the last relation. One should be in general  careful about the ordering when dealing with operators acting from the right.\\
\\
The reader may be tired by now of getting so much advice - so to get it a chance to be appreciated, let us start.
\\
\\
As for alternative form of the velocity operator in (\ref{V_i}), here we provide a sketch of the relevant derivation:

\begin{eqnarray}
\hat{V}_i &=& -i \left[\hat{x}_i, \hat{H}\right] \nonumber \\
 &=& -i \left[\hat{x}_i, \frac{1}{2\lambda r}(\hat{a}^+_\alpha \hat{a}_\alpha + \hat{b}^+_\alpha \hat{b}_\alpha - \hat{a}^+_\alpha \hat{b}_\alpha -  \hat{a}_\alpha \hat{b}^+_\alpha        ) -\frac{q}{\hat{r}}\right] \nonumber \\
  &=&  \frac{-i}{2\lambda \hat{r}}\left[\hat{x}_i, (\hat{a}^+_\alpha \hat{a}_\alpha + \hat{b}^+_\alpha \hat{b}_\alpha - \hat{a}^+_\alpha \hat{b}_\alpha -  \hat{a}_\alpha \hat{b}^+_\alpha ) \right] \nonumber \\
&=&  \frac{i}{4\hat{r}}\sigma^i_{\gamma\delta}\left[\hat{a}^+_\gamma \hat{a}_\delta + \hat{b}_\delta \hat{b}^+_\gamma \, , \, \hat{a}^+_\alpha \hat{b}_\alpha +  \hat{a}_\alpha \hat{b}^+_\alpha  \right] = \frac{i}{2 \hat{r}}\sigma^i_{\gamma\delta}( \hat{a}^+_\gamma \hat{b}_\delta -  \hat{a}_\delta \hat{b}^+_\gamma ) \nonumber \\
\label{Vcalc}&=& \frac{i}{2\hat{r}} \hat{w}_i
\end{eqnarray}

Calculations related to various forms in which $A_k$ can be written (the derivation of the equation (\ref{A_k})) involve:

\begin{eqnarray}
\hat{A}_k &= &\frac{1}{2}\varepsilon_{ijk} (\hat{L}_i\hat{V}_j + \hat{V}_j\hat{L}_i) + q \frac{\hat{x}_k}{\hat{r}} \,\, = \,\, \hat{A}_k^0 + q \frac{\hat{x}_k}{\hat{r}}
\end{eqnarray}

\begin{eqnarray}
\hat{A}^0_k &=& \varepsilon_{ijk} \frac{i}{8\hat{r}} \sigma^i_{\alpha \beta} \sigma^j_{\gamma \delta}  \nonumber \\
&& \times ((\hat{a}^+_\alpha \hat{a}_\beta - \hat{b}_\beta \hat{b}^+_\alpha)(\hat{a}^+_\gamma \hat{b}_\delta - \hat{a}_\delta \hat{b}^+_\gamma) +(\hat{a}^+_\gamma \hat{b}_\delta-\hat{a}_\delta \hat{b}^+_\gamma)(\hat{a}^+_\alpha \hat{a}_\beta - \hat{b}_\beta \hat{b}^+_\alpha))\nonumber\\
&& \nonumber \\
 &=& -\frac{1}{8\hat{r}} (\sigma^k_{\alpha \delta} \delta_{\gamma\beta} - \sigma^k_{\gamma \beta} \delta_{\alpha \delta}) \nonumber \\
  && \times ((\hat{a}^+_\alpha \hat{a}_\beta - \hat{b}_\beta \hat{b}^+_\alpha)(\hat{a}^+_\gamma \hat{b}_\delta - \hat{a}_\delta \hat{b}^+_\gamma)+(\hat{a}^+_\gamma \hat{b}_\delta-\hat{a}_\delta \hat{b}^+_\gamma)(\hat{a}^+_\alpha \hat{a}_\beta - \hat{b}_\beta \hat{b}^+_\alpha))\nonumber\\
&& \nonumber \\
 &=& -\frac{\sigma^k_{\alpha \delta}}{8\hat{r}} ((\hat{a}^+_\alpha \hat{a}_\beta - \hat{b}_\beta \hat{b}^+_\alpha)(\hat{a}^+_\beta \hat{b}_\delta - \hat{a}_\delta \hat{b}^+_\beta)+ (\hat{a}^+_\beta \hat{b}_\delta-\hat{a}_\delta \hat{b}^+_\beta)(\hat{a}^+_\alpha \hat{a}_\beta - \hat{b}_\beta \hat{b}^+_\alpha)) \nonumber\\
 && +\frac{ \sigma^k_{\gamma \beta}}{8\hat{r}}   ((\hat{a}^+_\alpha \hat{a}_\beta - \hat{b}_\beta \hat{b}^+_\alpha)(\hat{a}^+_\gamma \hat{b}_\alpha - \hat{a}_\alpha \hat{b}^+_\gamma)+(\hat{a}^+_\gamma \hat{b}_\alpha-\hat{a}_\alpha \hat{b}^+_\gamma)(\hat{a}^+_\alpha \hat{a}_\beta - \hat{b}_\beta \hat{b}^+_\alpha))\nonumber\\
&& \nonumber \\
&=& -\frac{\sigma^k_{\alpha \delta}}{8\hat{r}} (\hat{a}_\beta \hat{a}^+_\beta \hat{a}^+_\alpha \hat{b}_\delta  - \hat{a}^+_\alpha \hat{b}_\delta -\hat{a}^+_\alpha \hat{a}_\delta  \hat{a}_\beta \hat{b}^+_\beta -  \hat{b}^+_\alpha \hat{b}_\delta \hat{b}_\beta \hat{a}^+_\beta + \hat{a}^+_\alpha \hat{b}_\delta + \hat{b}_\beta \hat{b}^+_\beta \hat{b}^+_\alpha \hat{a}_\delta   )  \nonumber \\
&& -\frac{\sigma^k_{\alpha \delta}}{8\hat{r}}(  \hat{a}^+_\beta \hat{a}_\beta \hat{b}_\delta \hat{a}^+_\alpha - \hat{a}^+_\alpha \hat{b}_\delta  - \hat{b}_\delta \hat{b}^+_\alpha \hat{a}^+_\beta \hat{b}_\beta + \hat{a}^+_\alpha \hat{b}_\delta - \hat{a}_\delta \hat{a}^+_\alpha \hat{b}^+_\beta \hat{a}_\beta + \hat{b}^+_\beta \hat{b}_\beta \hat{a}_\delta \hat{b}^+_\alpha)  \nonumber \\
&& + \frac{ \sigma^k_{\gamma \beta}}{8\hat{r}}(\hat{a}_\beta \hat{a}^+_\gamma \hat{a}^+_\alpha \hat{b}_\alpha - \hat{a}^+_\gamma \hat{b}_\beta  - \hat{a}^+_\alpha \hat{a}_\alpha \hat{a}_\beta \hat{b}^+_\gamma - \hat{b}^+_\alpha \hat{b}_\alpha \hat{b}_\beta \hat{a}^+_\gamma + \hat{a}^+_\gamma \hat{b}_\beta + \hat{b}_\beta \hat{b}^+_\gamma \hat{b}^+_\alpha \hat{a}_\alpha  )   \nonumber \\
&& + \frac{ \sigma^k_{\gamma \beta}}{8\hat{r}}( \hat{a}^+_\gamma \hat{a}_\beta  \hat{b}_\alpha \hat{a}^+_\alpha     - \hat{a}^+_\gamma \hat{b}_\beta  -\hat{b}_\alpha \hat{b}^+_\alpha \hat{a}^+_\gamma \hat{b}_\beta  + \hat{a}^+_\gamma \hat{b}_\beta - \hat{a}_\alpha \hat{a}^+_\alpha \hat{b}^+_\gamma \hat{a}_\beta + \hat{b}^+_\gamma \hat{b}_\beta \hat{a}_\alpha \hat{b}^+_\alpha )   \nonumber \\
&& \nonumber \\
&=& -\frac{\sigma^k_{\alpha \delta}}{8\lambda \hat{r}}( (2\hat{r}^L + 2\hat{r}^R)\hat{a}^+_\alpha \hat{b}_\delta + (2\hat{r}^R + 2\hat{r}^L) \hat{b}^+_\alpha \hat{a}_\delta )  \nonumber \\
&& + \frac{1}{8\lambda \hat{r}} ( (2\hat{x}_k^L + 2\hat{x}_k^R) \hat{a}_\beta \hat{b}^+_\beta  + (2\hat{x}_k^R + 2\hat{x}_k^L) \hat{a}^+_\beta \hat{b}_\beta )              \nonumber \\
&& \nonumber \\
&=& -\frac{1}{2\lambda \hat{r}}\hat{r} \sigma^k_{\alpha \delta}(\hat{a}^+_\alpha \hat{b}_\delta +  \hat{b}^+_\alpha \hat{a}_\delta) + \frac{1}{2\lambda \hat{r}}\hat{x}_k (\hat{a}_\beta \hat{b}^+_\beta  +  \hat{a}^+_\beta \hat{b}_\beta )  \nonumber \\
\label{Acalc} &=&-\frac{1}{2\lambda \hat{r}}(\hat{r}\hat{\zeta}_k - \hat{x}_k \hat{\zeta})
\end{eqnarray}

\be\nonumber
\boxed{\hat{A}_k^0 =-\frac{1}{2\lambda \hat{r}}(\hat{r}\hat{\zeta} _k - \hat{x}_k \hat{\zeta})}
\ee

Now we are going to derive a few interesting identities which will come in handy when dealing with calculations related to the vanishing commutator $[\hat{W}', \hat{W}'_k]$ in  (\ref{dotA}) (recall  that for states with equal number of creation/annihilation operators $\hat{r}=\hat{r}^L=\hat{r}^R$).

\begin{eqnarray}
[\hat{\zeta}_k , \hat{r}^L] &=&     \lambda \sigma^k_{\alpha \beta} [ \hat{a}^+_\alpha \hat{b}_\beta + \hat{a}_\beta \hat{b}^+_\alpha , \hat{a}^+_\delta \hat{a}_\delta +1]
= \lambda \sigma^k_{\alpha \beta} (\hat{a}^+_\delta \hat{b}_\beta (-\delta_{\alpha\delta}) +\delta_{\beta\delta} \, \hat{a}_\delta \hat{b}^+_\alpha)     \nonumber  \\
&=&- \lambda \sigma^k_{\alpha \beta} (\hat{a}^+_\alpha \hat{b}_\beta - \hat{a}_\beta \hat{b}^+_\alpha)=  -\lambda \hat{w}_k =i 2\lambda \hat{r} \hat{V}_k    \nonumber
\end{eqnarray}

For the sake of symmetry and the next calculations let us also calculate

\begin{eqnarray}
[\hat{\zeta} , \hat{x}_k] &=& \frac{\lambda}{2}\sigma^k_{\alpha \beta} [\hat{a}^+_\delta \hat{b}_\delta + \hat{a}_\delta \hat{b}^+_\delta, \hat{a}^+_\alpha \hat{a}_\beta + \hat{b}_\beta \hat{b}^+_\alpha ]\nonumber  \\
&=&\frac{\lambda}{2} \sigma^k_{\alpha \beta}(\hat{a}^+_\alpha (- \delta_{\delta \beta}) \hat{b}_\delta + \hat{a}^+_\delta (-\delta_{\delta\alpha})\hat{b}_\beta + \hat{b}^+_\delta(\delta_{\delta \alpha}) \hat{a}_\beta + \hat{a}_\delta \hat{b}^+_\alpha ( \delta_{\delta \beta}) )\nonumber\\
&=&-\lambda \sigma^k_{\alpha \beta} (\hat{a}^+_\alpha \hat{b}_\beta - \hat{a}_\beta \hat{b}^+_\alpha)=-\lambda \hat{w}_k= i 2\lambda \hat{r} \hat{V}_k\nonumber
\end{eqnarray}

\be
\boxed{[\hat{\zeta} , \hat{x}_k]= [\hat{\zeta} _k , \hat{r}]=i 2\lambda \hat{r} \hat{V}_k}
\ee
\vspace*{0.3cm}

\begin{eqnarray}
[\hat{\zeta} , \hat{\zeta} _{\alpha \beta}]&=& [\hat{a}^+_\gamma \hat{b}_\gamma + \hat{a}_\gamma \hat{b}^+_\gamma\,,\,\hat{a}^+_\alpha \hat{b}_\beta + \hat{a}_\beta \hat{b}^+_\alpha ]\nonumber\\
&=&-\delta_{\gamma\beta}\,\hat{b}_\gamma \hat{b}^+_\alpha -\delta_{\gamma\alpha}\,\hat{a}_\beta \hat{a}^+_\gamma + \delta_{\gamma\alpha} \, \hat{b}^+_\gamma \hat{b}_\beta + \delta_{\gamma\beta}\,\hat{a}^+_\alpha \hat{a}_\gamma  \nonumber \\
&=& -\hat{b}_\beta \hat{b}^+_\alpha -\hat{a}_\beta \hat{a}^+_\alpha + \hat{b}^+_\alpha \hat{b}_\beta + \hat{a}^+_\alpha \hat{a}_\beta =\delta_{\alpha\beta} - \delta_{\alpha\beta} =0\nonumber
\end{eqnarray}
\be
  \boxed{[\hat{\zeta}, \hat{\zeta} _k]=0}
\ee
\vspace*{1cm}

At this point we have all that is needed to prove (\ref{dotA}):

\begin{eqnarray}
[\hat{W}', \hat{W}'_k] &=& [\hat{W} +\omega \hat{r}, \hat{W}_k + \omega \hat{x}_k]\nonumber \\
&=& \left[2\hat{r}\lambda^{-1} - \hat{\zeta} + \omega \hat{r}, 2\hat{x}_k\lambda^{-1} - \hat{\zeta} _k + \omega \hat{x}_k\right]\nonumber \\
&=& \eta^2 [ \hat{r}, \hat{x}_k] -  \eta [\hat{r},\hat{\zeta}_k] -\eta [\hat{\zeta} , \hat{x}_k]+[\hat{\zeta} ,\hat{\zeta}_k] \nonumber \\
&=&  0+ \eta 2i\hat{r}\lambda \hat{V}_k- \eta 2i\hat{r}\lambda  \hat{V}_k +0 = 0 \nonumber
\end{eqnarray}

\be\label{W0WK}
\boxed{[\hat{W}', \hat{W}'_k]=0}
\ee
\vspace*{1cm}

When checking the $so(4)$ algebra relations in (\ref{AiAj}),the knowledge of the commutator $[\hat{W}'_i, \hat{W}'_j]$ is necessary. In order to work it out, we need to derive also $[\hat{\zeta}_i,\hat{\zeta}_j],\, [\hat{x}_i, \hat{\zeta} _j ],\,[\hat{x}_i,\hat{x}_j] $. As to the following few calculations, recall that $\hat{x}_i$ is composed of the operators $x_i$ acting from the left and the right side: $\hat{x}_i=(\hat{x}_i^L + \hat{x}_i^R)/2$. Also it is suitable to consult the beginning of the Appendix, namely various identities which hold for the $\sigma$-matrices. Without that what follows may seem incomprehensible. For instance in the very next calculation of  $[\hat{\zeta}_i,\hat{\zeta}_j] $, writing down the second-to-last line after the previous one would be almost a blasphemy in the non-commutative world, if it were not for the fact that $\sigma$-matrices are traceless, i.e. $\sigma^k_{\alpha\alpha}=0$.

\begin{eqnarray}
[\hat{\zeta}_i,\hat{\zeta}_j] &=& \sigma^i_{\alpha \beta}\sigma^j _{\gamma \delta}  [\hat{\zeta}_{\alpha \beta} , \hat{\zeta} _{\gamma \delta}]=\sigma^i_{\alpha \beta}\sigma^j_{\gamma \delta} [\hat{a}^+_\alpha \hat{b}_\beta + \hat{a}_\beta \hat{b}^+_\alpha \,,\,\hat{a}^+_\gamma \hat{b}_\delta + \hat{a}_\delta \hat{b}^+_\gamma ]\nonumber\\
&=&-(\sigma^j\sigma^i)_{\gamma\beta} \hat{b}_\beta \hat{b}^+_\gamma - (\sigma^i\sigma^j)_{\alpha\delta } \hat{a}_\delta \hat{a}^+_\alpha + (\sigma^i\sigma^j)_{\alpha\delta} \hat{b}^+_\alpha \hat{b}_\delta + (\sigma^j\sigma^i)_{\gamma\beta} \hat{a}^+_\gamma \hat{a}_\beta
\nonumber\\
&=& \delta_{ij}(\hat{b}^+_\alpha \hat{b}_\alpha -  \hat{b}_\alpha \hat{b}^+_\alpha +    \hat{a}^+_\alpha \hat{a}_\alpha -  \hat{a}_\alpha \hat{a}^+_\alpha )+ \nonumber\\
 && + i \varepsilon_{ijk}\sigma^k_{\alpha\beta}(\hat{b}_\beta \hat{b}^+_\alpha + \hat{b}^+_\alpha \hat{b}_\beta - \hat{a}_\beta \hat{a}^+_\alpha - \hat{a}^+_\alpha \hat{a}_\beta)\nonumber \\
&=& \delta_{ij}(\delta_{\alpha\alpha}-\delta_{\alpha\alpha})+2i\varepsilon_{ijk}\sigma^k_{\alpha\beta}(\hat{b}_\beta \hat{b}^+_\alpha  - \hat{a}^+_\alpha \hat{a}_\beta)\nonumber \\
& =& 2i\lambda^{-1}\varepsilon_{ijk}(\hat{x}_k^R  - \hat{x}_k^L) = -4i\varepsilon_{ijk} \hat{L}_k \nonumber
\end{eqnarray}

\be
\boxed{[\hat{\zeta}_i,\hat{\zeta}_j]=-4i \varepsilon_{ijk} \hat{L}_k}
\ee
\vspace*{1cm}

\begin{eqnarray}
[\hat{x}_i^L, \hat{\zeta}_j] &=& \lambda \sigma^i _{\alpha \beta} \sigma^j _{\gamma \delta} [ \hat{a}^+_\alpha \hat{a}_\beta , \hat{a}^+_\gamma \hat{b}_\delta + \hat{a}_\delta \hat{b}^+_\gamma] \nonumber \\
&=&\lambda  \sigma ^i _{\alpha \beta} \sigma^j _{\gamma \delta}  (\hat{a}^+_\alpha \hat{b}_\delta \delta_{\beta \gamma} - \hat{a}_\beta \hat{b}^+_\gamma \delta_{\alpha \delta} \nonumber \\
& = &\lambda( (\sigma ^i \sigma ^j)_{\alpha \beta}\hat{a}^+_\alpha \hat{b}_\beta -(\sigma^j \sigma^i)_{\alpha \beta} \hat{a}_\beta \hat{b}^+_\alpha)\nonumber
\end{eqnarray}

\begin{eqnarray}
[\hat{x}_i^R, \hat{\zeta} _j] &=& \lambda \sigma ^i _{\alpha \beta} \sigma^j _{\gamma \delta} [ \hat{b}_\beta \hat{b}^+_\alpha  , \hat{a}^+_\gamma \hat{b}_\delta + \hat{a}_\delta \hat{b}^+_\gamma] \nonumber \\
&=& \lambda  \sigma ^i _{\alpha \beta} \sigma^j _{\gamma \delta}  (\hat{b}_\beta \hat{a}^+_\gamma \delta_{\alpha \delta} - \hat{a}_\delta \hat{b}^+_\alpha \delta_{\beta \gamma} )\nonumber \\
&=& \lambda (-(\sigma ^i \sigma ^j)_{\alpha \beta}\hat{a}_\beta \hat{b}^+_\alpha + (\sigma^j \sigma^i)_{\alpha \beta}\hat{a}^+_\alpha \hat{b}_\beta) \nonumber
\end{eqnarray}

\begin{eqnarray}
[\hat{x}_i, \hat{\zeta} _j ] &=& \frac{\lambda}{2}(\sigma ^i \sigma ^j+\sigma^j \sigma^i)_{\alpha \beta}(\hat{a}^+_\alpha \hat{b}_\beta - \hat{a}_\beta \hat{b}^+_\alpha) \nonumber \\
&=& \frac{\lambda}{2}\{\sigma^i, \sigma^j\} _{\alpha \beta} \hat{w}_{\alpha \beta} = \lambda\delta_{ij} \hat{w}\nonumber
\end{eqnarray}

\begin{equation}
\boxed{[\hat{x}_i,\hat{\zeta} _j] = \lambda \delta_{ij} \hat{w}}
\end{equation}
\vspace*{1cm}

The following calculation takes use of the fact that $\hat{x}_i^L$ commutes with $\hat{x}_i^R$. It is also useful to have in mind
$[\hat{x}_i^L ,\hat{x}_j^L]= 2i\varepsilon_{ijk}\lambda \hat{x}_k^L$ and $[\hat{x}_i^R ,\hat{x}_j^R]=  2i\varepsilon_{jik}\lambda \hat{x}_k^R$

\begin{eqnarray}
[\hat{x}_i,\hat{x}_j] &=& \frac{1}{4}[\hat{x}_i^L + \hat{x}_i^R, \hat{x}_j^L+ \hat{x}_j^R] =\frac{1}{4}([\hat{x}_i^L , \hat{x}_j^L] +[ \hat{x}_i^R,  \hat{x}_j^R] ) \nonumber \\
&=&     \frac{1}{4}(2i\varepsilon_{ijk}\lambda \hat{x}_k^L + 2i\varepsilon_{jik}\lambda \hat{x}_k^R)=  \frac{i}{2}\varepsilon_{ijk}(\hat{x}_k^L - \hat{x}_k^R)=  i \lambda^2 \varepsilon_{ijk}\hat{L}_k         \nonumber
\end{eqnarray}

\be
\boxed{[\hat{x}_i,\hat{x}_j]=i \lambda^2 \varepsilon_{ijk} \hat{L}_k}
\ee

So for the commutators involved in (\ref{AiAj}) we can write:

\begin{eqnarray}
 [\hat{W}'_i, \hat{W}'_j] &=& [\eta \hat{x}_i - \hat{\zeta} _i , \eta \hat{x}_j - \hat{\zeta} _j]\nonumber \\
&=& \eta^2 [\hat{x}_i, \hat{x}_j] - \eta ([\hat{x}_i , \hat{\zeta} _j] - [\hat{x}_j, \hat{\zeta} _i]) +[\hat{\zeta}_i , \hat{\zeta} _j]\nonumber\\
&=&\eta^2 i \varepsilon_{ijk} \lambda ^2 \hat{L}_k - \eta((\delta_{ij} -\delta_{ji})\lambda w) - i4 \varepsilon_{ijk} \hat{L}_k \nonumber \\
&=& 4i\lambda\omega\left(1+ \left(\frac{\lambda\omega}{4}\right)\right)\varepsilon_{ijk}\hat{L}_k  \nonumber
\end{eqnarray}

\be\label{tildWW}
\boxed{ [\hat{W}'_i, \hat{W}'_j]=4i\lambda\omega\left(1+ \left(\frac{\lambda\omega}{4}\right)\right)\varepsilon_{ijk}\hat{L}_k }
\ee

\vspace*{1cm}

To compute the commutator $[\hat{L}_i,\hat{A}_j]$  acting on the vectors from $\mathcal{H_{\lambda}^E}$, we actually need  to find $[\hat{L}_i, \hat{W}'_j]$ and multiply it with some constants. To manage this, let us start with

\begin{eqnarray}
[\hat{L}_i ,\hat{x}_j]&=& \frac{1}{4\lambda}[\hat{x}_i^L- \hat{x}_i^R, \hat{x}_j^L+ \hat{x}_j^R] =\frac{1}{4\lambda}([\hat{x}_i^L, \hat{x}_j^L]- [ \hat{x}_i^R,  \hat{x}_j^R] ) \nonumber\\
&=& \frac{1}{4\lambda}(2i\varepsilon_{ijk}\lambda \hat{x}_k^L - 2i\varepsilon_{jik}\lambda \hat{x}_k^R)=  \frac{i}{2}\varepsilon_{ijk}(\hat{x}_k^L +\hat{x}_k^R)= i \varepsilon_{ijk}\hat{x}_k \nonumber
\end{eqnarray}

\be\label{L_ix_j}
\boxed{ [\hat{L}_i ,\hat{x}_j]= i \varepsilon_{ijk}\hat{x}_k}
\ee

Somewhere above we have already computed $[\hat{x}_i^L, \hat{\zeta}_j]$ and  $[\hat{x}_i^R, \hat{\zeta}_j]$, we can use it also now:

\begin{eqnarray}
[\hat{L}_i , \hat{\zeta}_j]&=& \frac{1}{2\lambda}[\hat{x}_i^L- \hat{x}_i^R , \hat{\zeta}_j]=\frac{1}{2\lambda}([\hat{x}_i^L , \hat{\zeta}_j] -[ \hat{x}_i^R , \hat{\zeta}_j])\nonumber \\
&=& \frac{1}{2\lambda}\lambda((\sigma^i\sigma^j)_{\alpha\beta}(\hat{a}^+_\alpha \hat{b}_\beta + \hat{a}_\beta \hat{b}^+_\alpha )-(\sigma^j\sigma^i)_{\alpha\beta}(\hat{a}^+_\alpha \hat{b}_\beta + \hat{a}_\beta \hat{b}^+_\alpha )) \nonumber\\
&=&\frac{1}{2}[\sigma^i,\,\sigma^j]_{\alpha\beta}(\hat{a}^+_\alpha \hat{b}_\beta + \hat{a}_\beta \hat{b}^+_\alpha )=i\varepsilon_{ijk} \sigma^k_{\alpha\beta}(\hat{a}^+_\alpha \hat{b}_\beta + \hat{a}_\beta \hat{b}^+_\alpha ) = i\varepsilon_{ijk}\hat{\zeta}_k \nonumber
\end{eqnarray}

\be\label{L_izeta_j}
\boxed{ [\hat{L}_i , \hat{\zeta}_j]=i \varepsilon_{ijk} \hat{\zeta}_k  }
\ee

\begin{eqnarray}\label{liwj}
[\hat{L}_i , \hat{W}'_j]&=& [\hat{L}_i , \eta \hat{x}_j - \hat{\zeta}_j]=- i \varepsilon_{ijk} (\eta \hat{x}_k - \hat{\zeta}_k )\nonumber \\
&=&   i \varepsilon_{ijk} \hat{W}'_k
\end{eqnarray}

At this point to obtain the commutator $[\hat{L}_i , \hat{A}_j]$ in $(\ref{LA})$ we just need to recall that $\frac{\hat{W}'_j}{2\lambda}= \hat{A}_j$, and therefore
\be
\boxed{ [\hat{L}_i , \hat{A}_j]=i \varepsilon_{ijk} \hat{A}_k }
\ee

\vspace*{1cm}
Here we provide some details related to the first Casimir operator in (\ref{CAS}). We are going to prove that  $\hat{L}_j \hat{A}_j \propto \hat{L}_j (\eta \hat{x}_k - \hat{\zeta}_j)=0 $ when restricted to the subspace $\mathcal{H}_{\lambda}^E$. In the very next calculation a lot of renaming of the dummy indices appears and the reader will probably appreciate having a piece of paper and a pencil at hand in order to keep track of the procedure. The final zero in the last line is obtained after realizing that there is no difference between $\hat{r}^L$ and  $\hat{r}^R$ if they act on anything which has the same number of creation and annihilation  operators. Our considered operator wave functions $\Psi$ have this property and the same holds for $\hat{a}^+_\alpha \hat{b}_\alpha \Psi$ and $\hat{a}_\alpha \hat{b}^+_\alpha \Psi$.

\begin{eqnarray}
2 \hat{L}_j \hat{\zeta} _j &=& \sigma ^j _{\alpha \beta} \sigma^j_{\gamma \delta}  (\hat{a}^+_\alpha \hat{a}_\beta - \hat{b}_\beta \hat{b}^+_\alpha )(\hat{a}^+_\gamma \hat{b}_\delta + \hat{a}_\delta \hat{b}^+\gamma) \nonumber \\
 &=& (2 \delta_{\alpha\delta}\delta_{\beta\gamma}-\delta_{\alpha\beta}\delta_{\gamma\delta})  (\hat{a}^+_\alpha \hat{a}_\beta - \hat{b}_\beta \hat{b}^+_\alpha )(\hat{a}^+_\gamma \hat{b}_\delta + \hat{a}_\delta \hat{b}^+\gamma) \nonumber \\
&=& 2 ( \hat{a}^+_\alpha \hat{a}_\beta \hat{a}^+_\beta \hat{b}_\alpha + \hat{a}^+_\alpha \hat{a}_\beta \hat{a}_\alpha \hat{b}^+_\beta -  \hat{b}_\beta \hat{b}^+_\alpha \hat{a}^+_\beta \hat{b}_\alpha - \hat{b}_\beta \hat{b}^+_\alpha \hat{a}_\alpha \hat{b}^+_\beta) \nonumber \\
&& - ( \hat{a}^+_\alpha \hat{a}_\alpha \hat{a}^+_\beta \hat{b}_\beta + \hat{a}^+_\alpha \hat{a}_\alpha \hat{a}_\beta \hat{b}^+_\beta -  \hat{b}_\alpha \hat{b}^+_\alpha \hat{a}^+_\beta \hat{b}_\beta -  \hat{b}_\alpha \hat{b}^+_\alpha \hat{a}_\beta \hat{b}^+_\beta) \nonumber \\
&=& 2(\hat{a}_\beta \hat{a}^+_\beta \hat{a}^+_\alpha \hat{b}_\alpha - \hat{a}^+_\alpha \hat{b}_\alpha + \hat{a}^+_\alpha \hat{a}_\alpha \hat{a}_\beta \hat{b}^+_\beta -\hat{b}^+_\alpha \hat{b}_\alpha \hat{b}_\beta \hat{a}^+_\beta + \hat{a}^+_\alpha \hat{b}_\alpha -\hat{b}_\beta \hat{b}^+_\beta \hat{b}^+_\alpha \hat{a}_\alpha )  \nonumber \\
&& -\hat{a}_\beta \hat{a}^+_\beta \hat{a}^+_\alpha \hat{b}_\alpha +2 \hat{a}^+_\alpha \hat{b}_\alpha - \hat{a}^+_\alpha \hat{a}_\alpha \hat{a}_\beta \hat{b}^+_\beta + \hat{b}^+_\alpha \hat{b}_\alpha \hat{b}_\beta \hat{a}^+_\beta -2\hat{a}^+_\alpha \hat{b}_\alpha \nonumber \\
&\,& + \hat{b}_\beta \hat{b}^+_\beta \hat{b}^+_\alpha \hat{a}_\alpha  \nonumber \\
&=& \hat{a}_\beta \hat{a}^+_\beta \hat{a}^+_\alpha \hat{b}_\alpha + \hat{a}^+_\beta \hat{a}_\beta \hat{a}_\alpha \hat{b}^+_\alpha - \hat{b}^+_\beta \hat{b}_\beta \hat{a}^+_\alpha \hat{b}_\alpha - \hat{b}_\beta \hat{b}^+_\beta \hat{a}_\alpha \hat{b}^+_\alpha \nonumber \\
&=& \lambda^{-1}\left( (\hat{r}^L + \lambda)-(\hat{r}^R + \lambda) \right) \hat{a}^+_\alpha \hat{b}_\alpha + \lambda^{-1}\left( (\hat{r}^L - \lambda)-(\hat{r}^R - \lambda) \right) \hat{a}_\alpha \hat{b}^+_\alpha \nonumber \\ &=& \,\,0 \nonumber
\end{eqnarray}
\vspace*{0.5cm}
\begin{equation}\label{zeroCas1}
\boxed{\hat{L}_j \hat{\zeta}_j=0}
\end{equation}

\vspace*{1cm}

\begin{eqnarray}
\hat{L}_j \hat{x} _j &=& \frac{1}{4\lambda} (\hat{x}_j^L - \hat{x}_j^R)(\hat{x}_j^L + \hat{x}_j^R) = \frac{1}{4\lambda}(\hat{x}_j^L \hat{x}_j^L + \hat{x}_j^L  \hat{x}_j^R - \hat{x}_j^R  \hat{x}_j^L - \hat{x}_j^R  \hat{x}_j^R )  \nonumber \\
&=& \frac{1}{4\lambda}(\hat{x}_j^L \hat{x}_j^L  - \hat{x}_j^R  \hat{x}_j^R ) =  \frac{1}{2\lambda}(((\hat{r}^L)^2-\lambda^2) - ((\hat{r}^2)^R - \lambda^2) )= 0 \nonumber
\end{eqnarray}
\begin{equation}\label{zeroCas2}
\boxed{\hat{L}_j \hat{x}_j=0}
\end{equation}

 \vspace*{1cm}

Putting the previous two calculations together we can claim
\be
\hat{L}_j \hat{A}_j = \frac{\hat{L}_j \hat{W}'_j}{2 \lambda}  = \frac{\hat{L}_j (\eta \hat{x}_j -\hat{\zeta}_j)}{2 \lambda} = 0.
\ee

\vspace*{1cm}

Now let us move to the second Casimir operator in (\ref{CAS}). We have mentioned already that the relevant calculation is based on the fact which we are about to prove now:

\be \label{id}
 \hat{W}'_i\hat{W}'_i +(\eta^2\lambda^2-4)(\hat{L}_i\hat{L}_i+1) = \hat{W}'^2
\ee

 Undoubtedly it is a useful identity, but the reader may be slightly uncertain about where did the very idea come from.
  Perhaps the following few words  will fail to  give  a satisfactory motivation, but it will have to suffice:
  There are not too many equations which relate operators and c-numbers, so we naturally try to use them whenever possible. If we encounter a term which seems someway related to the Schr\"{o}dinger equation,  we are only glad to work towards rewriting that expression in terms of $\hat{W}'$. (Just to refresh the reader's memory, one form of the Schr\"{o}dinger equation reads $(\hat{W}' - 2 \lambda q) \Psi_E= 0$.)
  As to the noticing the very fact that the expression on the left hand side  of (\ref{id}) is indeed the case - a suspicion may arise when expanding it suitably - but one can hardly avoid using some  trial and error method anyway.

So let us start. Calculations of many identities follow. The  motivation for some of them may not be  straightforward, but all should be clear by the end of the paragraph at the latest.

\begin{eqnarray}
 \hat{W}'^2 &=&  (\eta \hat{r} - \hat{\zeta})^2 = \eta^2 \hat{r}^2 -\eta \{\hat{r}, \hat{\zeta} \} + (\hat{\zeta} )^2 \nonumber \\
 & &    \nonumber \\
 \hat{W}'_i \hat{W}'_i&=&  (\eta \hat{x}_i - \hat{\zeta}_i)^2 = \eta^2 \hat{x}_i \hat{x}_i -\eta \{\hat{x}_i, \hat{\zeta}_i\} + \hat{\zeta}_i \hat{\zeta}_i
\end{eqnarray}

\begin{eqnarray}
 \hat{L}_i \hat{L}_i &=&  \frac{1}{4\lambda^2}( \hat{x}^L_i - \hat{x}^R_i)^2 = \frac{1}{4\lambda^2}( (r^L)^2 + (\hat{r}^R)^2 - 2\hat{x}^L_i\hat{x}^R_i - 2\lambda^2) \nonumber
 \end{eqnarray}

 \be
\boxed{ \hat{L}_i \hat{L}_i = \frac{1}{2\lambda^2}( \hat{r}^2 - \hat{x}^L_i\hat{x}^R_i - \lambda^2) }
\ee

\begin{eqnarray}
 \hat{x}_i \hat{x}_i &=&  \frac{1}{4}( \hat{x}^L_i + \hat{x}^R_i)^2 = \frac{1}{4}( (\hat{r}^L)^2 + (\hat{r}^R)^2 + 2\hat{x}^L_i\hat{x}^R_i - 2\lambda^2) \nonumber
 \end{eqnarray}
  \be
\boxed{ \hat{x}_i \hat{x}_i = \frac{1}{2}( \hat{r}^2 + \hat{x}^L_i\hat{x}^R_i - \lambda^2) }
\ee

\begin{eqnarray}
 \lambda^{-2}\hat{x}^L_i \hat{x}^R_i &=& \sigma^i_{\alpha\beta}\sigma^i_{\gamma\delta}(\hat{a}^+_\alpha \hat{a}_\beta \hat{b}_\delta \hat{b}^+_\gamma)= (2\delta_{\alpha\delta}\delta_{\gamma\beta}-\delta_{\alpha\beta}\delta_{\gamma\delta})(\hat{a}^+_\alpha \hat{a}_\beta \hat{b}_\delta \hat{b}^+_\gamma)\nonumber \\
 &=&   2 \hat{a}^+_\alpha \hat{a}_\beta \hat{b}_\alpha \hat{b}^+_\beta -     \hat{a}^+_\alpha \hat{a}_\alpha \hat{b}_\beta \hat{b}^+_\beta =    2 \hat{a}^+_\alpha \hat{a}_\beta \hat{b}_\alpha \hat{b}^+_\beta -     \lambda^{-2}(\hat{r}^L-\lambda)(\hat{r}^R-\lambda) \nonumber
 \end{eqnarray}
  \be
\boxed{ \lambda^{-2}\hat{x}^L_i\hat{x}^R_i =  2 \hat{a}^+_\alpha \hat{a}_\beta \hat{b}_\alpha \hat{b}^+_\beta - \lambda^{-2}(\hat{r}^2 -2\lambda \hat{r} + \lambda^2 )}
\ee

The last result is to be used to rewrite the $2\hat{a}^+_\alpha \hat{a}_\beta \hat{b}_\alpha \hat{b}^+_\beta $ term in the following calculation:

\begin{eqnarray}
( \hat{\zeta} )^2 &=& (\hat{a}^+_\alpha \hat{b}_\alpha + \hat{a}_\alpha \hat{b}^+_\alpha )(\hat{a}^+_\beta \hat{b}_\beta + \hat{a}_\beta \hat{b}^+_\beta )\nonumber \\
 &=& \hat{a}^+_\alpha \hat{a}^+_\beta \hat{b}_\alpha \hat{b}_\beta + \hat{a}_\alpha \hat{a}_\beta \hat{b}^+_\alpha \hat{b}^+_\beta +       \hat{a}^+_\alpha \hat{a}_\beta \hat{b}_\alpha \hat{b}^+_\beta  + (\hat{a}^+_\beta \hat{a}_\alpha +\delta_{\alpha\beta} )(\hat{b}_\beta \hat{b}^+_\alpha +\delta_{\alpha\beta})                         \nonumber \\
&=&  \hat{a}^+_\alpha \hat{a}^+_\beta \hat{b}_\alpha \hat{b}_\beta + \hat{a}_\alpha \hat{a}_\beta \hat{b}^+_\alpha \hat{b}^+_\beta + 2 \hat{a}^+_\alpha \hat{a}_\beta \hat{b}_\alpha \hat{b}^+_\beta + 2\lambda^{-1}\hat{r} \nonumber \\
&=&  \hat{a}^+_\alpha \hat{a}^+_\beta \hat{b}_\alpha \hat{b}_\beta + \hat{a}_\alpha \hat{a}_\beta \hat{b}^+_\alpha \hat{b}^+_\beta + \lambda^{-2}(\hat{x}^L_i \hat{x}^R_i + \hat{r}^2 + \lambda^2) \nonumber
 \end{eqnarray}

   \be
\boxed{ ( \hat{\zeta} )^2 = \hat{a}^+_\alpha \hat{a}^+_\beta \hat{b}_\alpha \hat{b}_\beta + \hat{a}_\alpha \hat{a}_\beta \hat{b}^+_\alpha \hat{b}^+_\beta + \lambda^{-2}(\hat{x}^L_i \hat{x}^R_i + \hat{r}^2 + \lambda^2)}
\ee
The above equation will be used to rewrite the terms $\hat{a}^+_\alpha \hat{a}^+_\beta \hat{b}_\alpha \hat{b}_\beta + \hat{a}_\alpha \hat{a}_\beta \hat{b}^+_\alpha \hat{b}^+_\beta$ in the following one:

\begin{eqnarray}
 \hat{\zeta}^{i}  \hat{\zeta}^{i} &=&   \sigma^i_{\alpha\beta}\sigma^i_{\gamma\delta}\hat{\zeta}_{\alpha\beta}\hat{\zeta}_{\gamma\delta}= (2\delta_{\alpha\delta}\delta_{\gamma\beta}-\delta_{\alpha\beta}\delta_{\gamma\delta})(\hat{a}^+_\alpha \hat{b}_\beta + \hat{a}_\beta \hat{b}^+_\alpha)(\hat{a}^+_\gamma \hat{b}_\delta + \hat{a}_\delta \hat{b}^+_\gamma)  \nonumber \\
 &=&  2 (\hat{a}^+_\alpha \hat{a}^+_\beta \hat{b}_\alpha \hat{b}_\beta + \hat{a}^+_\alpha \hat{a}_\alpha \hat{b}_\beta \hat{b}^+_\beta + \hat{a}_\beta \hat{a}^+_\beta \hat{b}^+_\alpha \hat{b}_\alpha + \hat{a}_\alpha \hat{a}_\beta \hat{b}^+_\alpha \hat{b}^+_\beta ) -    ( \hat{\zeta}^{0})^2                 \nonumber \\
&=&  2 (\hat{a}^+_\alpha \hat{a}^+_\beta \hat{b}_\alpha \hat{b}_\beta +  \hat{a}_\alpha \hat{a}_\beta \hat{b}^+_\alpha \hat{b}^+_\beta ) +4\lambda^{-2}(\hat{r}^2 + \lambda^2) -    ( \hat{\zeta} )^2                         \nonumber \\
&=&  ( \hat{\zeta} )^2  + 2\lambda^{-2}(\hat{r}^2 - \hat{x}_i^L \hat{x}_i^R +\lambda^2)                       \nonumber
 \end{eqnarray}

  \be
\boxed{ \hat{\zeta}^{i}  \hat{\zeta}^{i} =  ( \hat{\zeta} )^2  + 2\lambda^{-2}(\hat{r}^2 - \hat{x}_i^L\hat{x}_i^R +\lambda^2)}
\ee

\begin{eqnarray}
[\hat{r}, \hat{\zeta} ] &=& \frac{\lambda}{2}[\hat{a}^+_\alpha \hat{a}_\alpha + \hat{b}_\alpha \hat{b}^+_\alpha + 2,\, \hat{a}^+_\beta \hat{b}_\beta + \hat{a}_\beta \hat{b}^+_\beta]   \nonumber \\
 &=& \lambda(\hat{a}^+_\alpha \hat{b}_\alpha - \hat{a}_\alpha \hat{b}^+_\alpha)  = \lambda \hat{w} \nonumber \\
 & & \nonumber \\
 \{\hat{r}, \hat{\zeta} \} &=& [\hat{r}, \hat{\zeta} ] + 2 \hat{\zeta}  \hat{r} =  \lambda \hat{w} + 2 \hat{\zeta}  \hat{r} \nonumber
 \end{eqnarray}

  \be
\boxed{  [\hat{r}, \hat{\zeta} ]  = \lambda \hat{w}   \ \ \ \ \ \ \ \    \{\hat{r}, \hat{\zeta} \}  =  \lambda \hat{w} + 2 \hat{\zeta}  \hat{r}  }
\ee

\begin{eqnarray}
\hat{\zeta}_i \hat{x}_i &=&  \frac{\lambda}{2}\sigma^i_{\alpha\beta}\sigma^i_{\gamma\delta}( \hat{a}^+_\alpha \hat{b}_\beta + \hat{a}_\beta \hat{b}^+_\alpha) ( \hat{a}^+_\gamma \hat{a}_\delta + \hat{b}_\delta \hat{b}^+_\gamma) \nonumber \\
&=& \frac{\lambda}{2}(2\delta_{\alpha\delta}\delta_{\gamma\beta}-\delta_{\alpha\beta}\delta_{\gamma\delta}) ( \hat{a}^+_\alpha \hat{b}_\beta + \hat{a}_\beta \hat{b}^+_\alpha) ( \hat{a}^+_\gamma \hat{a}_\delta + \hat{b}_\delta \hat{b}^+_\gamma) \nonumber \\
 &=&  \lambda (\hat{a}^+_\alpha \hat{a}_\alpha \hat{a}^+_\beta \hat{b}_\beta -\hat{a}^+_\alpha \hat{b}_\alpha  + \hat{b}_\beta \hat{b}^+_\beta \hat{a}^+_\alpha \hat{b}_\alpha -\hat{a}^+_\alpha \hat{b}_\alpha + \hat{a}_\beta \hat{a}^+_\beta \hat{a}_\alpha \hat{b}^+_\alpha + \hat{b}^+_\alpha \hat{b}_\alpha \hat{a}_\beta \hat{b}^+_\beta  )  \nonumber \\
 & & - \frac{1}{2}\hat{\zeta} (\hat{r}^L + \hat{r}^R - 2\lambda)\nonumber \\
 & = &  (\hat{r}^L + \hat{r}^R - 2\lambda) \hat{a}^+_\alpha \hat{b}_\alpha +  (\hat{r}^L + \hat{r}^R + 2\lambda) \hat{a}_\alpha \hat{b}^+_\alpha -2\lambda \hat{a}^+_\alpha \hat{b}_\alpha - \hat{\zeta}  \hat{r} + \lambda \hat{\zeta}        \nonumber \\
  & =& \hat{r}\hat{\zeta}  + [\hat{r},\,\hat{\zeta} ] - 3\lambda \hat{w} \ =\ \hat{r}\hat{\zeta}  - 2\lambda \hat{w} \nonumber
 \end{eqnarray}

 \begin{eqnarray}
\hat{x}_i\hat{\zeta}_i  &=&  \frac{\lambda}{2}\sigma^i_{\alpha\beta}\sigma^i_{\gamma\delta}( \hat{a}^+_\alpha \hat{a}_\beta + \hat{b}_\beta \hat{b}^+_\alpha) ( \hat{a}^+_\gamma \hat{b}_\delta + \hat{a}_\delta \hat{b}^+_\gamma) \nonumber \\
&=&  \frac{\lambda}{2}(2\delta_{\alpha\delta}\delta_{\gamma\beta}-\delta_{\alpha\beta}\delta_{\gamma\delta})( \hat{a}^+_\alpha \hat{a}_\beta + \hat{b}_\beta \hat{b}^+_\alpha) ( \hat{a}^+_\gamma \hat{b}_\delta + \hat{a}_\delta \hat{b}^+_\gamma) \nonumber \\
 &=&  \lambda (\hat{a}_\beta \hat{a}^+_\beta \hat{a}^+_\alpha \hat{b}_\alpha -\hat{a}^+_\alpha \hat{b}_\alpha  + \hat{a}^+_\alpha \hat{a}_\alpha \hat{a}_\beta \hat{b}^+_\beta + \hat{b}^+_\alpha \hat{b}_\alpha \hat{a}^+_\beta \hat{b}_\beta   -\hat{a}^+_\alpha \hat{b}_\alpha  + \hat{b}_\beta \hat{b}^+_\beta \hat{a}_\alpha \hat{b}^+_\alpha  )  \nonumber \\
 & & - \frac{1}{2}(\hat{r}^L + \hat{r}^R - 2\lambda)\hat{\zeta}  \nonumber \\
 & = &  (\hat{r}^L + \hat{r}^R + 2\lambda) \hat{a}^+_\alpha \hat{b}_\alpha +  (\hat{r}^L + \hat{r}^R - 2\lambda) \hat{a}_\alpha \hat{b}^+_\alpha -2\lambda \hat{a}^+_\alpha \hat{b}_\alpha - \hat{r}\hat{\zeta}  + \lambda \hat{\zeta}       \nonumber \\
  & =& \hat{r}\hat{\zeta}  +\lambda \hat{w}  \nonumber
 \end{eqnarray}

  \be
\boxed{  [\hat{x}_i,\, \hat{\zeta}_i]  = 3\lambda \hat{w}   \ \ \ \ \ \ \ \    \{  \hat{x}_i,\, \hat{\zeta}_i \}  =  2 \hat{r}\hat{\zeta}  - \lambda \hat{w}    }
\ee

Now we can use all the derived stuff to prove (\ref{id}):

 \begin{eqnarray}
 (\hat{W}' )^2 &=&  \eta^2 \hat{r}^2 -\eta \{\hat{r}, \hat{\zeta} \} + (\hat{\zeta} )^2 =   \eta^2 \hat{r}^2 - \eta (\lambda \hat{w} + 2\hat{\zeta}  \hat{r}) + (\hat{\zeta} )^2 \nonumber \\
 & & \nonumber \\
 & & \nonumber \\
  (\hat{W}'_i)^2 &+& (\eta^2\lambda^2 -4 )((\hat{L}_i)^2 + 1) = \nonumber \\
 & &  \nonumber \\
 &=&  \frac{\eta^2}{2}(\hat{r}^2 + \hat{x}_i^L \hat{x}_i^R - \lambda^2) -\eta (2\hat{r}\hat{\zeta} -\lambda w) + (\hat{\zeta} )^2 + \frac{2}{\lambda^2}(\hat{r}^2 - \hat{x}_i^L \hat{x}_i^R + \lambda^2)   \nonumber \\
  & & + \frac{\eta^2}{2}(\hat{r}^2 - x_i^L x_i^R - \lambda^2) - \frac{2}{\lambda^2}(r^2 - x_i^L x_i^R - \lambda^2) + \eta^2\lambda^2 -4 \nonumber \\
 & = & \eta^2 \hat{r}^2 - \eta(2\hat{r}\hat{\zeta}  - \lambda \hat{w}) + (\hat{\zeta} )^2  \nonumber \\
 &=& (\hat{W}')^2 \nonumber
  \end{eqnarray}

In the last line we have used $[\hat{r}, \hat{\zeta} ]=\lambda \hat{w}$ and consequently
\be
 \lambda \hat{w} + 2\hat{\zeta}  \hat{r} =  2\hat{r}\hat{\zeta}  - \lambda \hat{w}
  \ee

The proof of    (\ref{id}) is finished, and  (\ref{CAS2}) follows immediatedly.

\bibliography{aipsamp}%
 
\end{document}